\documentclass[notitlepage,article,onecolumn,pre,aps,floatfix]{revtex4-1}
\usepackage{amsmath}
\usepackage{amsthm}
\usepackage{amssymb}
\usepackage{mathrsfs}
\usepackage{graphicx}
\usepackage{epstopdf}
\usepackage[usenames]{color}
\usepackage{xr}
\usepackage{placeins}
\usepackage[hidelinks]{hyperref}
\usepackage[normalem]{ulem}
\usepackage{rotating}
\usepackage{verbatim}
\usepackage{bibentry}
\usepackage{xspace}
\usepackage[margin=1.0in]{geometry}
\usepackage{bm}
\usepackage{amssymb}
\usepackage{soul}
\usepackage{hyperref}
\usepackage{siunitx}
\usepackage{floatrow}
\usepackage[toc,page]{appendix}

\theoremstyle{plain}
\newtheorem*{proposition*}{Diffusion approximation}

\pdfoutput=1

\begin{document}

\title{Steady states of active Brownian particles interacting with boundaries}

\author{Caleb G. Wagner}
\email{c.g.wagner23@gmail.com}
\affiliation{Mechanical and Materials Engineering, University of Nebraska-Lincoln, Lincoln, NE 68588}
\author{Michael F. Hagan}
\email{hagan@brandeis.edu}
\author{Aparna Baskaran}
\email{aparna@brandeis.edu}

\affiliation{Martin Fisher School of Physics, Brandeis University, Waltham, Massachusetts 02453, USA.}

\begin{abstract}
An active Brownian particle is a minimal model for a self-propelled colloid in a dissipative environment. Experiments and simulations show that, in the presence of boundaries and obstacles, active Brownian particle systems approach nontrivial nonequilibrium steady states with intriguing phenomenology, such as accumulation at boundaries, ratchet effects, and long-range depletion interactions. Nevertheless, theoretical analysis of these phenomena has proven difficult. Here we address this theoretical challenge in the context of non-interacting particles in two dimensions, basing our analysis on the steady-state Smoluchowski equation for the 1-particle distribution function. Our primary result is an approximation strategy that connects asymptotic solutions of the Smoluchowski equation to boundary conditions. We test this approximation against the exact analytic solution in a 2d planar geometry as well as numerical solutions in circular and elliptic geometries. We find good agreement so long as the boundary conditions do not vary too rapidly with respect to the persistence length of particle trajectories. Our results are relevant for characterizing long-range flows and depletion interactions in such systems. In particular, our framework shows how such behaviors are connected to the breaking of detailed balance at the boundaries.
\end{abstract}

\maketitle

\section{Introduction} \label{sec:introduction}

Active matter describes a class of systems that are maintained far from equilibrium by driving forces acting on the constituent particles \cite{Ramaswamy2010,Marchetti2013,Bechinger2016,Zottl2016,Yoshinaga2017,Ramaswamy2017,Needleman2017,Saintillan2018,Fodor2018,Seifert2019}. Experimental realizations span many scales, from the microscopic and colloidal to the macroscopic. Examples on the microscopic level are reconstituted biopolymers and molecular motors \cite{Sanchez2012}, bacterial suspensions \cite{Dombrowski2004,Kaiser2014}, and synthetic self-propelled colloids \cite{Chate2012,Bartolo2013,Palacci2013a}. On larger scales, many macrobiological systems can be thought of as active matter, including swarming midges \cite{vanderVaart2019,Cepelewicz2019}, aggregating fire ants \cite{Tennenbaum2015}, schooling fish \cite{Becco2006,Cambui2012}, and flocking birds \cite{Attanasi2014b}. There are also examples of macroscopic active systems constructed artificially, such as vibrated self-propelled granular particles \cite{Narayan2006a,Narayan2007,Walsh2016,Dauchot2017,Walsh2017,Scholz2018} and small robots \cite{Savoie2018,Hartnett2018,Deblais2018,Dauchot2019}. 

Active matter is a promising testing bed for the development of new ideas about nonequilibrium steady states in general. It is already known that active matter steady states exhibit many unusual properties. Examples are the nonexistence of state variables like pressure and temperature \cite{Solon2015b,Dauchot2017}, violations of extensivity \cite{Wagner2017}, aggregation at boundaries \cite{Wensink2008,Volpe2011,Elgeti2013,Lee2013,Ezhilan2015}, and athermal phase separation \cite{Fily2012,Redner2013a,Stenhammar2013,Stenhammar2014,Wysocki2014,Cates2015,Stenhammar2015,Ni2014,Redner2016}. Moreover, while equilibrium systems obey detailed balance and cannot exhibit net currents, this is no longer true out of equilibrium. Strikingly, in active matter systems, currents can occur even without external driving, instead resulting from rectification of the particle-level active driving by way of spatial asymmetries in the boundary conditions \cite{Wan2008,Reichhardt2017}.

At the concrete level, progress toward a fundamental understanding of these unusual phenomena requires minimal models that reproduce characteristically nonequilibrium phenomenology while still being amenable to theoretical treatment. In this work, we focus on a simple, well-studied model for active matter, namely, a non-interacting collection of active Brownian particles (ABPs). This minimal model exhibits several of the features of nonequilibrium steady states mentioned above, including accumulation at boundaries, nonlocal probability measures, and long-ranged currents in steady-state. In particular, the existence of long-ranged currents was first shown by Baek, et al \cite{Solon2018}, who derived a multipole expansion for the density and current exterior to a fixed, passive inclusion. These solutions do not require external forcing, in agreement with previous simulation results on active ratchet systems \cite{Reichhardt2017}. Instead, geometric constraints imposed by boundaries rectify the active driving, inducing currents. As an example, the current exterior to a fixed, elliptical inclusion asymptotically decays as $r^{-2}$, and the density as $r^{-1}$.

While Ref. \cite{Solon2018} has shown that long-range, current-carrying solutions do exist, it is not obvious how to connect these solutions with boundary conditions. In this work, we seek to further understanding of ABP steady states by developing an approximation scheme that connects such asymptotic (long-range) behavior of the steady state with boundary conditions. We demonstrate the utility of this scheme by considering planar, circular, and elliptic geometries.  In particular, we identify the mechanism by which certain surface-particle interactions generate long-range dipolar and quadrupolar flow fields. We verify our analytical results using two numerical methods: the first, a trajectory-sampling technique based on the stochastic particle dynamics and the second, the finite element software PDE2D \cite{Note1}.

Our framework proves that long-range currents are tied with the breaking of detailed balance at the boundaries. We argue that this connection has far-reaching implications for nonequilibrium steady states in general, not just those in active matter. Further, our work provides insight into pathways to construct finite current steady states via the design of boundary conditions. As we provide intuitive and computationally cheap methods to connect boundary conditions to material properties in bulk, they can be used to significantly constrain the space of design possibilities to obtain such steady states. One can then harness these types of long-range density variations and spontaneous flows to engineer active baths with tunable functionality, such as ``depletion-induced'' long-ranged forces between passive inclusions that can be either attractive or repulsive depending on the configuration and design \cite{Angelani2011,Ray2014,Harder2014,Ni2015,Leite2016,ZaeifiYamchi2017,Solon2018,Li2020,Zarif2020,Torrik2021}.

\section{Model definitions} \label{sec:model_definitions}

\begin{figure}
  \includegraphics[keepaspectratio=true,scale=0.32]{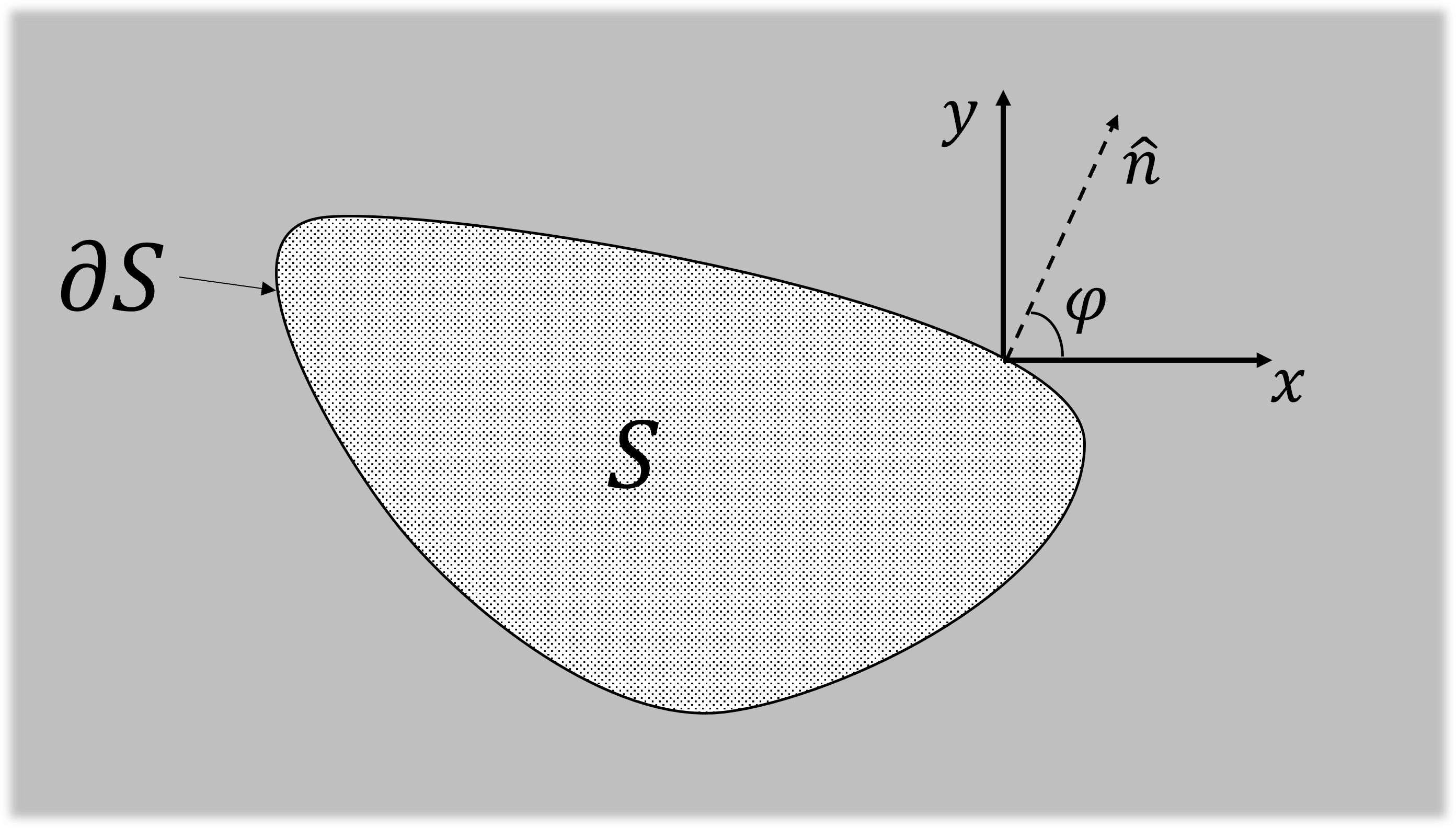}
  \caption{Illustration of the problems of interest in this work. The region $S$ models a fixed inclusion with hard walls. The boundary $\partial S$ is an idealized approximation of the narrow region $\mathbb{B}$ on which $\nabla V(\mathbf{r})$ is steep and repulsive. Outside of $S$, $\nabla V(\mathbf{r})$ is zero and the ABPs move freely with a uniform density boundary condition at infinity.}
  \label{fig:2d-domain}
\end{figure}

 We consider active Brownian particles in 2d. The dynamics of the center of mass $\mathbf{r}(t) = (x(t), y(t))$ and $\hat{u} = (\cos \theta, \sin \theta)$, the direction of the self-propulsion velocity, is given by
\begin{align}
 \dot{\mathbf{r}} &= -\xi^{-1} \nabla V(\mathbf{r}) +  v_0 \hat{u} + \sqrt{2 D_t} \boldsymbol{\eta}^\text{T}  \label{eq:introduction-1} \\
 \dot{\theta} &= \sqrt{2 D_r} \eta^\text{R}  \label{eq:introduction-2}
\end{align}
Here $\xi$ is the friction (units of [mass]/[time]), $V(\mathbf{r})$ is an external potential, $v_0$ is the magnitude of the self-propulsion velocity, and $D_t$ and $D_r$ are the translational and rotational diffusion coefficients.  The $\eta$ variables are Gaussian white noise with $\langle
\eta_i(t)\rangle = 0$ and $\langle \eta_i(t) \eta_j(t')\rangle =\delta_{ij} \delta(t - t')$. In this paper, we study the steady-state statistics of Eqs. \eqref{eq:introduction-1}--\eqref{eq:introduction-2} in terms of the associated Fokker-Planck equation for the probability density function $f(\mathbf{r}, \theta)$:
\begin{equation}
\ell \,  \hat{u} \cdot \nabla f = -(\xi D_r)^{-1} \nabla \cdot \left[\nabla V(\mathbf{r}) f \right] + (D_t / D_r) \nabla^2 f + \partial^2_{\theta} f \label{eq:introduction-3}
\end{equation}
where $\ell \equiv v_0 / D_r$ is a length associated with the persistence of particle trajectories in space. Previous work has used this equation as a starting point for various approximate descriptions of ABP steady states. Nevertheless, solutions which are both exact and nontrivial are essentially nonexistent. In this work we first approximate the Fokker-Planck equation itself, isolating the effects of activity by neglecting thermal noise (setting $D_t = 0$). In addition, we let $\nabla V(\mathbf{r})$ simulate an impenetrable, passive inclusion occupying a bounded region $S$ of $\mathbb{R}^2$ (Fig. \ref{fig:2d-domain}). Specifically, we make $\nabla V(\mathbf{r})$ large on a narrow region defining the boundary of $S$ (denoted $\partial S$), and $0$ otherwise. Then, in the region exterior to $S$, \eqref{eq:introduction-3} becomes
\begin{equation}
\ell \, \hat{u} \cdot \nabla f  = \partial^2_{\theta} f  \label{eq:introduction-4}
 \end{equation}
 Mathematically, an equation of this form requires so-called ``half-range'' boundary conditions on $\partial S$, which \emph{only} specify the distribution of particles pointing \emph{outward} from $\partial S$ \cite{Greenberg1987}. That is, if $\hat{n}$ is the outward unit normal vector at a point $\mathbf{r}_0$ on $\partial S$, then $f(\mathbf{r}_0, \theta)$ is specified only where $\hat{n} \cdot \hat{u} > 0$. Physically, these boundary conditions derive from the details of the particle-boundary interactions, which occur just within $\partial S$. For instance, if $\nabla V(\mathbf{r})$ is steep and repulsive, then, to a good approximation, particles leave the boundary as soon as their self-propulsion velocity points away from it, which causes $f(\mathbf{r}_0, \theta)$ to have strong peaks near $\hat{n} \cdot \hat{u} \approx 0$ and be $0$ otherwise \cite{Wagner2017}.

\section{Review of 1d problems} \label{sec:review_of_1d_problems}

Our framework for 2d problems is motivated in part by our previous treatment of 1d problems \cite{Wagner2017,Wagner2019a}, for which Eq. \eqref{eq:introduction-4} simplifies to
\begin{equation}
\ell \cos \theta \frac{\partial f(x, \theta)}{\partial x}=\frac{\partial ^{2}f(x, \theta)}{\partial \theta ^{2}} \label{eq:1dproblem-1}.
\end{equation}
Here, we consider this equation on a finite domain $0 < x < W$ and with boundary conditions
\begin{align}
&f(0,\theta ) =v_{+}(\theta ),\text{ \ \ \ \ \ where } \cos(\theta) >0 \label{eq:1dproblem-2a} \\
&f(W,\theta ) =v_{-}(\theta ),\text{ \ \ \ \ where } \cos(\theta) <0 \label{eq:1dproblem-2b}
\end{align}
for given $v_{+}(\theta )$. Note that Eqs. \eqref{eq:1dproblem-2a} and \eqref{eq:1dproblem-2b} conform with the type of boundary condition from section \ref{sec:model_definitions}, with $v_{\pm}(\theta )$ being derived from the interactions of the ABPs with walls at $x = 0, W$.

Here, we solve this problem by separation of variables. The separable solutions take the form $e^{-r_k (x/\ell)} \phi_k(\theta)$, where
\begin{equation}
\frac{d^2 \phi_k}{d \theta^2} + r_k \cos \theta \phi_k = 0 \label{eq:1dproblem-3}
\end{equation}
Together with periodic boundary conditions in $\theta$, this equation is an eigenvalue problem of the indefinite Sturm-Liouville type. It can be shown that there are an infinite number of eigenvalues which are discrete, real, and anti-symmetric about $0$; we index these as
\begin{equation}
\ldots r_{-2} < r_{-1} < r_0 < r_{1} < r_{2} < \ldots \label{eq:1dproblem-4}
\end{equation}
where $r_0 = 0$. Refs. \onlinecite{Wagner2017} and \onlinecite{Wagner2019a} contain further details on $r_k$ and $\phi_k$ and their computation.

The usual strategy is to expand the general solution in terms of the separable solutions. However, for the problem \eqref{eq:1dproblem-1}--\eqref{eq:1dproblem-2b}, the separable solutions do not span the solution space. Instead, a non-separable solution $x - \ell \cos \theta$ is required, so that the general solution is
\begin{equation}
f(x, \theta) = c_0 + d_0 (x - \ell \cos \theta) + \sum_{k>0}a_{k}e^{-r _{k}(x/\ell)}\phi_k+\sum_{k<0}a_{k}e^{-r _{k}\left(x-W\right)/\ell }\phi_k.
\label{eq:1dproblem-5}
\end{equation}
The problem \eqref{eq:1dproblem-1}--\eqref{eq:1dproblem-2b} will then be solved if we can find $c_0$, $d_0$, and $a_k$ such that the boundary conditions \eqref{eq:1dproblem-2a}--\eqref{eq:1dproblem-2b} are satisfied. Specifically, this amounts to the requirement that, in addition to $c_0 + d_0 (x - \ell \cos \theta)$, the positive half of the spectrum generates a complete set over the interval where $\cos \theta > 0$, and similarly for the negative half of the spectrum on $\cos \theta < 0$. For the eigenvalue problem \eqref{eq:1dproblem-3}, this result indeed holds \cite{Wagner2019a}, which solves the problem at a formal level. Practical aspects of computing the expansion coefficients are addressed in Refs. \onlinecite{Wagner2017} and \onlinecite{Wagner2019a}. For present purposes, we note that the separable solutions in \eqref{eq:1dproblem-5} decay exponentially with distance from the boundary, so the behavior of Eq. \eqref{eq:introduction-1} in the bulk is dominated by the remainder $c_0 + d_0 (x - \ell \cos \theta)$. Therefore, if the constants $c_0$ and $d_0$ can be measured or inferred, then one obtains a significantly simplified picture of the steady-state statistics far from boundaries. Our goal in the next section is to replicate this simplification for 2d problems.

\section{Method of solution for 2d problems} \label{sec:method_of_solution}

Motivated by Eq. \eqref{eq:1dproblem-5}, we look for a way to decompose fully 2d problems into boundary layer and bulk (asymptotic) parts. The generic existence of a boundary layer is implied by the first-order spatial gradients in \eqref{eq:introduction-4}, which suggest solutions that decay exponentially over the length scale $\ell$, similarly to the separable solutions $e^{-r_k (x/\ell)} \phi_k(\theta)$ in the 1d problem. These solutions would be non-analytic in $\ell$, which raises the question of whether solutions exist which \emph{are} analytic in $\ell$, i.e., expressible as the power series
\begin{equation}
f \sim f_0 + \ell f_1 + \ell^2 f_2 + \ldots \label{eq:introduction-9}
\end{equation}
Such solutions are candidates for the bulk, or asymptotic, part of the solution. Indeed, they are known to exist for a broad class of integro-differential equations, called ``transport equations'', that are similar to \eqref{eq:introduction-4} in abstract structure and model various transport phenomena \cite{Greenberg1987,Beals1987,Cercignani1990,WagnerDissertation}. Because the asymptotic solutions \eqref{eq:introduction-9} persist into the bulk, they function as a useful starting point for coarse-grained models. For instance, in the Boltzmann equation for dilute gases, they generate classical hydrodynamics and its extensions, e.g., the Euler, Navier-Stokes, and Burnett equations \cite{Chapman1970,Grad1958,Cercignani1990}.

It is now apparent that the non-separable solution $x - \ell \cos \theta$ we introduced for 1d problems has the form of Eq. \eqref{eq:introduction-9}, and can be derived by substituting \eqref{eq:introduction-9} into Eq. \eqref{eq:introduction-1}. In fact, we can use the same method to construct the asymptotic solution for fully 2d problems. As explained below, this solution behaves like a diffusion process, so we denote it by $f_d$. Substituting \eqref{eq:introduction-9} into \eqref{eq:introduction-4} and matching powers of $\ell$ gives
\begin{equation}
f_d(x,y,\theta) = c_0 + \ell \, \beta(x,y) - \ell^2 \left( \frac{\partial \beta(x,y)}{\partial x} \cos \theta + \frac{\partial \beta(x,y)}{\partial y} \sin \theta \right) + \mathcal{O}(\ell^3) \label{eq:sec3-10}
\end{equation}
where $c_0$ is a constant, and $\beta(x,y)$ is an as-yet unknown function satisfying $\nabla^2 \beta = 0$. At quadrupole order, a coordinate-free representation is
\begin{equation}
f_d(\mathbf{r}, \theta) = c_0 + \ell \, \beta - \ell^2 \, \boldsymbol{\nabla} \beta \cdot \mathbf{\hat{u}} + \frac{\ell^3}{4} \left( \partial_{i} \partial_j \beta \right) \hat{u}_i \hat{u}_j + \ldots \label{eq:sec3-11}
\end{equation}
By integrating Eq. \eqref{eq:sec3-10} over $\theta$, we see that $\beta(x,y)$ is proportional to the density $\rho_d(x,y)$ (minus an additive constant) and so $\nabla^2 \rho_d = 0$. Moreover, integrating against $\hat{u}$ results in a Fickian expression for the particle flux $\mathbf{J}_d(x,y)$:
\begin{equation}
\mathbf{J}_d(x,y) = \int \hat{u} f_d(x,y, \theta) \, d\theta = -\frac{\ell^2}{2} \boldsymbol{\nabla} \rho_d(x,y) \label{eq:sec3-12}
\end{equation}
Therefore, the solution \eqref{eq:sec3-10} has characteristics of a diffusive process. For this reason, we call it the \emph{diffusion solution}.

As the asymptotic part of the solution, $f_d$ in principle provides an accurate description of ABP steady states far from boundaries. However, there is the significant caveat that the boundary conditions for $\beta(x,y)$ -- and by extension $f_d$ itself -- are not known \emph{a priori}. In general, it is not possible to obtain these without solving for the full distribution function, which includes the boundary layer. Addressing this challenge is one of the primary goals of this paper. Our starting point is the interesting observation that the separable solutions in the 1d problem are orthogonal to $\cos \theta$:
\begin{equation}
\int e^{-r_k (x/\ell)} \phi_k(\theta) \, \cos \theta \, d \theta = 0
\end{equation}
Combined with the expression for the general solution, Eq. \eqref{eq:1dproblem-5}, one obtains a simple, linear relation between the \emph{wall-normal flux} $J_{\mathrm{n}} = \int f(x,\theta) \cos \theta \, d \theta$ -- which is independent of $x$ -- and the parameters $c_0$ and $d_0$ of the bulk solution.  In particular, if $J_{\mathrm{n}}$ is known, then one can infer the bulk solution up to a normalization factor, without having to treat the boundary layer part of the solution directly.
The caveat is that $J_{\mathrm{n}}$ cannot be calculated in a simple way from the boundary data on the full PDE, because these are specified only on half of the $\theta$ domain. The values on the other half can only be obtained by solving the full problem, which includes the separable solutions in \eqref{eq:1dproblem-5}. On the other hand, $J_{\mathrm{n}}$ itself is a physical quantity that can be estimated or measured, in which case the bulk solution is a simplified and self-contained model of the steady state far from boundaries. 
Our strategy is to generalize this approach to fully 2d problems. There, the flux varies with $\mathbf{r}$, and so we assume, at a minimum, that the boundary-normal flux $J_{\text{n}}$ is known at each point on the boundary. Here, $\mathbf{J}$ is defined as $\int \hat{u} f(\mathbf{r}, \theta) \, d \theta$, so that
\begin{equation}
J_{\text{n}} \equiv \mathbf{J} \cdot \hat{n} = \int (\hat{u} \cdot \hat{n}) f(\mathbf{r}, \theta) \, d \theta
\end{equation}
where $\hat{n}$ is the outward pointing normal vector on the boundary. Ideally, the boundary layer part of the solution would be orthogonal to $\hat{u} \cdot \hat{n}$, in which case the asymptotic part $f_d$ would carry \emph{all} the wall-normal flux. Then, the unknown function $\beta(x,y)$ in the expression for $f_d$ (Eq. \eqref{eq:sec3-10}) would satisfy $\hat{n} \cdot \mathbf{J} = \hat{n} \cdot \nabla \beta$ on the boundary, i.e., acquire a Neumann boundary condition. Because $\beta(x,y)$ is also constrained to satisfy Laplace's equation $\nabla^2 \beta = 0$, knowing $\hat{n} \cdot \mathbf{J}$ would uniquely determine $\beta(x,y)$ (up to an additive constant) and thereby the entirety of the asymptotic solution, Eq. \eqref{eq:sec3-10}.
In reality, the situation is not so simple: as we show in section \ref{sec:planar-geometry}, the boundary layer contribution to the solution is no longer guaranteed orthogonal to $\hat{u} \cdot \hat{n}$. In the same section, nevertheless, we provide evidence that orthogonality does hold \emph{approximately} in situations where the boundary data do not vary too rapidly with $\ell$. We therefore build our approach using the following approximation:
\begin{proposition*}
The diffusion solution $f_d(\mathbf{r}, \theta)$ carries all the particle flux normal to the boundary, that is,
\begin{equation}
\int \left[ \left. f(\mathbf{r}, \theta) \right|_{\mathbf{r} \in \partial S'} - \left. f_d(\mathbf{r}, \theta) \right|_{\mathbf{r} \in \partial S'} \right] (\hat{u} \cdot \hat{n}) \, d\theta = 0; \quad \rightarrow \quad \quad \hat{n} \cdot \mathbf{J} =  \hat{n} \cdot \mathbf{J}_d \label{eq:sec3-13}
\end{equation}
where $\hat{n}$ is the unit normal vector at a point on the boundary, and $f(\mathbf{r}, \theta)$ is the full solution to the problem. $\mathbf{J}$ and $\mathbf{J}_d$ correspond respectively to the fluxes associated with the full solution and the diffusion solution.
\end{proposition*}
Note that this approximation says nothing about the flux component \emph{parallel} to the boundary. In fact, the diffusion solution for the 1d problem does not make any contribution to this component, so for 2d problems there is no rationale for an approximation like Eq. \eqref{eq:sec3-13} given in terms of the parallel flux. On the other hand, the diffusion solution \emph{does} give the complete flux vector far from the boundaries, from Eq. \eqref{eq:sec3-10} or \eqref{eq:sec3-12}. It is in this sense that Eq. \eqref{eq:sec3-13} connects long-range fluxes with the breaking of detailed balance at the boundaries, that is, a boundary normal flux that is not strictly $0$. Formally,
\begin{enumerate}
	\item A system breaks detailed balance at the boundary if $J_{\text{n}}$ is not strictly $0$.
	\item By Eq. \eqref{eq:sec3-13}, this leads to a nonzero Neumann boundary condition on $\beta(x,y)$.
	\item Because $\beta(x,y)$ itself satisfies Laplace's equation, it will be non-constant in such a case (and long-ranged via
its multipole expansion).
	\item By Eqs. \eqref{eq:sec3-10} and \eqref{eq:sec3-12}, the density and current will also be long-ranged in this sense.
\end{enumerate}

In the following sections, we work out this method of solution in detail for a selection of problems, corresponding to planar, circular, and elliptic geometries. Along with validating the diffusion approximation, we show that it is possible to obtain reasonable \emph{a priori} estimates of the boundary normal flux $\hat{n} \cdot \mathbf{J}$ in physically relevant situations. 

\section{Benchmarking the framework: planar geometry} \label{sec:planar-geometry}
As a first step, we validate our assertion that the diffusion approximation captures the bulk properties of the ABPs in the presence of boundaries. We do this by comparing the diffusion approximation with an exact solution to the planar 2d problem. Specifically, we consider a system with a wall at $x \mathop{=} 0$ and solve for the steady-state distribution $f(x,y,\theta)$ in the region $0 \mathop{<} x \mathop{<} \infty$ and $-\infty \mathop{<} y \mathop{<} \infty$ with boundary conditions 
\begin{align}
f(0, y, \theta) &= g(y, \theta), \qquad \,  \text{where } \cos \theta > 0 \\
f(x, y, \theta) &\rightarrow 0, \qquad \qquad \, \, \text{as } x \rightarrow \infty
\end{align}

To simplify notation, we work in units where $\ell = 1$.

\subsubsection{Separation of variables} \label{subsubsec:planar-geometry-b1}
In the Cartesian coordinates relevant for this problem, Eq.~\eqref{eq:introduction-4} is separable. Taking $f \mathop{=} \Gamma(x) \Xi(y) \Theta(\theta)$ leads to
\begin{align}
&\frac{d \Gamma}{dx} + \lambda \, \Gamma= 0; \qquad \frac{d \Xi}{dy} + \nu \, \Xi = 0 \label{eq:2d-planar-1}  \\
&\frac{d^2 \Theta}{d \theta^2} + (\lambda \cos \theta + \nu \sin \theta) \, \Theta = 0 \label{eq:2d-planar-3}
\end{align}
where $\lambda$ and $\nu$ are constants. To satisfy the general $y$-dependence of the boundary condition $x = 0$, the functions $\Xi(y)$ need to be complete on $-\infty < y < \infty$. Thus, we take $\nu$ to be purely imaginary but otherwise unconstrained, writing $\nu = \mu i$, where $\mu$ is real. We hope that for each $\mu$, Eqs. \eqref{eq:2d-planar-1}--\eqref{eq:2d-planar-3} admit a sequence of eigenvalues $\lambda_k(\mu)$ and eigenfunctions $\Theta_k(\theta, \mu)$ which are complete in $\theta$.
\subsubsection{Diffusion solutions} \label{subsubsec:planar-geometry-b2}
Let us now consider the asymptotic solution in the diffusion approximation formulated above,
\begin{equation}
f_d(x,y,\theta) = c_0 + \epsilon \, \beta(x,y) - \epsilon^2 \left( \cos \theta \frac{\partial \beta(x,y)}{\partial x}+ \sin \theta \frac{\partial \beta(x,y)}{\partial y} \right) + \mathcal{O}(\epsilon^3) \label{eq:2d-planar-4}
\end{equation}
where $\nabla^2 \beta(x,y) = 0$. We anticipate separable solutions with a $(x, y)$ dependence of the form $e^{c x} e^{\pm i c y}$, where $c$ is a constant. This suggests taking $\lambda = \pm \mu$ in Eqs. \eqref{eq:2d-planar-1}--\eqref{eq:2d-planar-3}. Eq. \eqref{eq:2d-planar-3} then becomes
\begin{equation}
\frac{d^2 \Theta}{d \theta^2} \pm \mu e^{\pm i \theta}  \Theta = 0 \label{eq:2d-planar-5}
\end{equation}
Making the change of variable $u = 2 i \, (\pm \mu)^{1/2} e^{\pm i \theta / 2}$ this turns into
\begin{equation}
u^2 \frac{d^2 \Theta}{d u^2} + u \frac{d\Theta}{du} + u^2 \Theta = 0 \label{eq:2d-planar-6}
\end{equation}
which is Bessel's equation with order $0$. The periodic boundary condition in $\theta$ implies $\Theta(u) = \Theta(-u)$, and thus the desired solution is the Bessel function of the first kind $J_0(u)$. In summary, the functions
\begin{equation}
f_d(x,y,\theta,\mu) = e^{\mp \mu x} e^{-\mu i y} J_0 \left(2 i \, (\pm \mu)^{1/2} e^{\pm i \theta / 2} \right) \label{eq:2d-planar-7}
\end{equation}
are solutions of \eqref{eq:introduction-4} for any real $\mu$. A direct calculation from \eqref{eq:sec3-11} shows that these are indeed the diffusion solutions.

\subsubsection{Boundary layer solutions} \label{subsubsec:planar-geometry-b3}

Now, to construct the exact solution, we need to tackle the boundary layer, and this is difficult to calculate. Let us begin by making the following change of variables,  $(\lambda, \mu ) = (r \cos \gamma, -i \, r \sin \gamma)$, where $\lambda$, $\mu$, $r$ are real and $\gamma$ is complex. Substitution into \eqref{eq:2d-planar-3} gives
\begin{equation}
\frac{d^2 \Theta}{d \omega^2} + r \cos \omega \, \Theta = 0 \label{eq:2d-planar-10}
\end{equation}
where $\omega = \theta - \gamma$ and this equation needs to be solved with periodic boundary conditions. Eq. \eqref{eq:2d-planar-10} has the same form as Eq. \eqref{eq:1dproblem-3} describing the angular eigenfunctions for 1d problems. We recall that there are an infinite number of eigenvalues which are discrete, real, and anti-symmetric about $0$, which we indexed with $k$ according to $\ldots r_{-2} < r_{-1} < r_0 < r_{1} < r_{2} < \ldots$, where $r_0 = 0$. Thus, the spectrum of the 2-parameter problem \eqref{eq:2d-planar-3} can be expressed as
\begin{align}
\lambda_k(\mu) &= \text{sgn}(k) \sqrt{|r_k|^2 + \mu^2}, \qquad k \in \mathbb Z \setminus \{0\} \label{eq:2d-planar-12} \\
\lambda^{\pm}_0(\mu) &= \pm |\mu| \label{eq:2d-planar-13}
\end{align}
We denote the corresponding eigenfunctions as $\Theta_k (\theta, \mu)$ and $\Theta^{\pm}_0 (\theta, \mu)$. Like the $\mu = 0$ case, the spectrum is discrete, real, and anti-symmetric about $0$. Nevertheless, there are important differences. While the eigenvalues $\lambda_k(\mu)$ are real, the eigenfunctions $\Theta_k(\theta, \mu)$ are not. In addition, $0$ is no longer an eigenvalue for nonzero $\mu$. Instead, the null space of the $\mu = 0$ problem transforms into the space spanned by $\Theta^{\pm}_0 (\theta, \mu)$, which are the diffusion solutions from the previous section. Explicitly,
\begin{align}
\Theta^{\pm}_0 (\theta, \mu) &= J_0 \left(2 i \, |\mu|^{1/2} e^{\pm\text{sgn}(\mu) i \theta / 2} \right) \label{eq:2d-planar-14}
\end{align}
such that the diffusion solutions are
\begin{equation}
f_d(x, y, \theta, \mu) = e^{\mp |\mu| x} e^{-\mu i y} \, \Theta_0^{\pm}(\theta, \mu) \label{eq:2d-planar-16}
\end{equation}

Using the change of variable from the previous section, the $\Theta_k(\theta, \mu)$ for $k \neq 0, \mu \neq 0$ can be constructed through analytic continuation of the known eigenfunctions for $\mu= 0$. Ref. \cite{Wagner2017} represents the latter as even/odd Fourier series:
{ \everymath={\displaystyle}
\begin{align}
\Theta_{k}(\omega, 0) = \left\{
\begin{array}{cc}
\frac{A^{(k)}_{0}(0)}{2} + \sum_{m = 1}^{\infty} A^{(k)}_{m}(0) \cos m \omega & \quad k \text{ even}  \\
 \sum_{m = 1}^{\infty}A^{(k)}_{m}(0) \sin m \omega & \quad  k \text{ odd}
\end{array}
\right. \label{eq:2d-planar-17}
\end{align}
}
The coefficients $A^{(k)}_{n}(0)$ are real and can be computed as eigenvectors of an infinite tridiagonal matrix \cite{Wagner2017}. We adopt the following normalization convention:
\begin{equation}
\int_{-\pi}^{\pi} \Theta_{k}(\omega, 0)^2 \cos \omega \, d \omega = \text{sgn}(k); \qquad \Theta_{k}(\omega + \pi, 0) = \Theta_{-k}(\omega, 0) \label{eq:2d-planar-18}
\end{equation}
which also fixes the normalization on $\Theta_{k}(\omega, \mu)$ for general $\mu$. Next, changing variable $\omega \rightarrow \theta - \gamma$ gives
{ \everymath={\displaystyle}
\begin{align}
\Theta_{k}(\theta - \gamma, 0) = \left\{
\begin{array}{cc}
 \frac{A^{(k)}_{0}(\mu)}{2} + \sum_{m = 2}^{\infty} A^{(k)}_{m}(\mu) \cos m \theta + \sum_{m = 2}^{\infty} B^{(k)}_{m}(\mu) \sin m \theta & \quad k \text{ even}  \\
-\sum_{m = 1}^{\infty} B^{(k)}_{m}(\mu) \cos m \theta + \sum_{m = 1}^{\infty} A^{(k)}_{m}(\mu) \sin m \theta & \quad  k \text{ odd}
\end{array}
\right. \label{eq:2d-planar-19}
\end{align} 
}
where
\begin{align}
A^{(k)}_{n}(\mu) = \cos \left[ n \arctan \left( \text{sgn}(k) \frac{i \mu}{\sqrt{|r_k|^2 + \mu^2}} \right) \right] \cdot A^{(k)}_{n}(0) \label{eq:2d-planar-20}  \\
B^{(k)}_{n}(\mu) = \sin \left[ n \arctan \left( \text{sgn}(k) \frac{i \mu}{\sqrt{|r_k|^2 + \mu^2}} \right) \right] \cdot A^{(k)}_{n}(0) \label{eq:2d-planar-21}
\end{align}
One can check by substitution that \eqref{eq:2d-planar-19} gives solutions of Eq. \eqref{eq:2d-planar-3}. So we make the association
\begin{equation}
\Theta_{k}(\theta - \gamma, 0) = \Theta_{k}(\theta, \mu) \label{eq:2d-planar-22}
\end{equation}

\subsubsection{Full solution of a boundary value problem}

To construct a general solution to the 2d planar problem, the separable solutions must satisfy some completeness relations in the $\theta$ variable. Towards this end, it is reasonable to expect that the $\Theta_{k}(\theta, \mu)$ for $\mu \neq 0$ satisfy full- and half-range completeness theorems analogous to those which hold for the $\mu = 0$ eigenfunctions. Numerical evidence suggests that such theorems hold for the real part of the $\Theta_k$, denoted by $\Omega_k$ \cite{WagnerDissertation}:
\begin{equation}
\Omega_k(\theta, |\mu|) \equiv \Re\left[\Theta_k(\theta, \mu)\right] \, \,= \frac{\Theta_{k}(\theta, \mu) + \Theta_{k}(\theta, -\mu)}{2} \label{eq:2d-planar-24}
\end{equation}
where we have used the fact that $\Theta_{k}(\theta, -\mu)$ is the complex conjugate of $\Theta_{k}(\theta, \mu)$. To formulate the completeness conjectures, let $\mathscr{H}$ be the Hilbert space associated with the inner product $\langle f, g \rangle = \int_{-\pi}^{\pi} f(\theta) g(\theta) |\cos \theta| \, d\theta$, and let $\mathscr{H}_{\pm}$ be the respective subspaces where $\cos \theta > 0$ or $\cos \theta < 0$. Then, we conjecture the following:
\begin{itemize}
\item{Full-range completeness ---} For real and non-zero $\mu$, any real-valued function in $\mathscr{H}$ can be expanded in terms of the functions $\left\{ \Omega_k(\theta, |\mu|),\Omega^{\pm}_0(\theta, |\mu|) \right\}$, where $k \in \mathbb Z \setminus \{0\}$.
\item{Half-range completeness ---} For real and non-zero $\mu$, any real-valued function in $\mathscr{H}_{+}$ can be expanded in terms of the functions $\left\{ \Omega_k(\theta, |\mu|),\Omega^{+}_0(\theta, |\mu|) \right\}$, where $k > 0$. Similarly, any real-valued function in $\mathscr{H}_{-}$ can be expanded in terms of $\left\{ \Omega_k(\theta, |\mu|),\Omega^{-}_0(\theta, |\mu|) \right\}$, where $k < 0$.
\end{itemize}

The imaginary part of $\Theta_k$ does not appear to share the completeness properties of the real part. As explained below and in appendix \ref{appendix:2d-problems}, however, both real and imaginary parts are required for solving the full boundary value problem. In anticipation of this, we define
\begin{equation}
\Upsilon_k(\theta, |\mu|) \equiv \Im\left[\Theta_k(\theta, \mu)\right] = \frac{\Theta_{k}(\theta, \mu) - \Theta_{k}(\theta, -\mu)}{2i} \label{eq:2d-planar-25}
\end{equation}

\begin{figure}
 \includegraphics[keepaspectratio=true,scale=0.36]{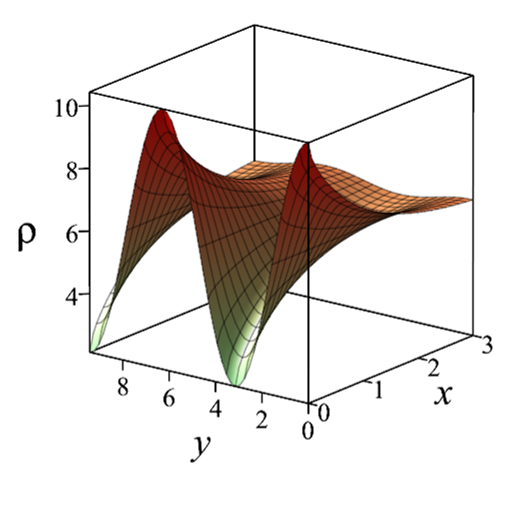}
 \includegraphics[keepaspectratio=true,scale=0.36]{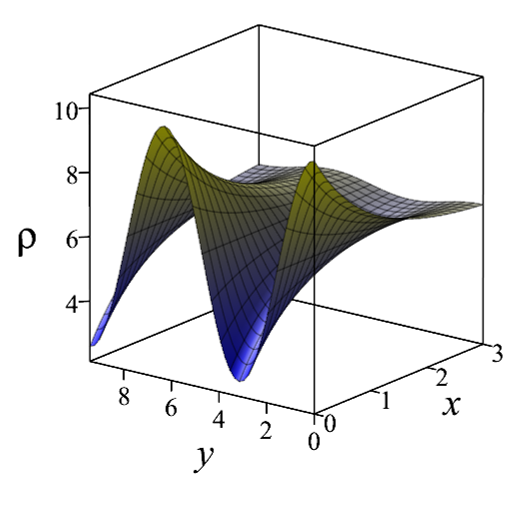}
  \includegraphics[keepaspectratio=true,scale=0.36]{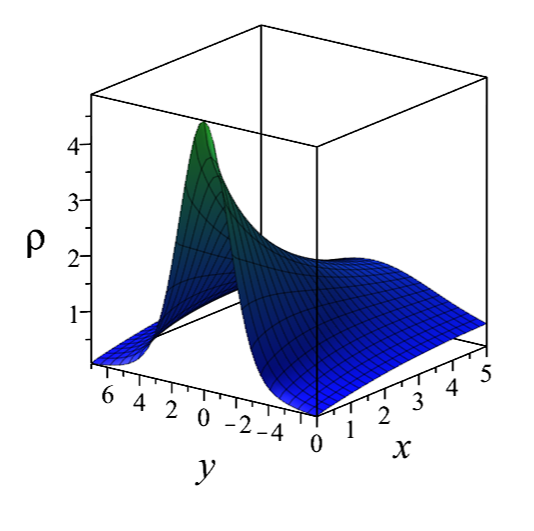} \hspace{5mm}
  \caption{Densities for the planar problem with two example boundary conditions. Left and middle: $f(0,y, \theta) \mathop{=} 2 + \cos y$, with the left displaying the density $\rho(x,y)$ computed using \eqref{temp343323543} and the least-squares method and the middle showing the density contribution from the diffusion solution, $\rho_d(x,y) \mathop{\simeq} 3.693 \times e^{-x} \,  \cos y$. The two qualitatively agree, validating the diffusion approximation. Right:  $f(0,y, \theta) \mathop{=} e^{-y^2/4}$, density computed using the least squares method. Comparing to the left panel, this illustrates the strong influence of the boundary on the bulk solution.}
  \label{fig:3d-planar-cos}
\end{figure}

Now, our task is to express $f(x, y, \theta)$ as a linear combination of separable solutions from above, varying $\lambda$ and $\mu$ over the range of allowed values. In view of the asymptotic condition on $f$, $\lambda$ must be positive, which restricts us to half of the spectrum. We then have
\begin{align}
f(x, y, \theta) = \int_{-\infty}^{\infty} d \mu \left[ \sum_{k \geq 0} a_k(\mu) e^{-\lambda_k(\mu) x} \Theta_k(\theta, \mu) \right] e^{-i \mu y} \label{eq:temp1543525676}
\end{align}
where the $k = 0$ mode corresponds to the sign of $\mu$, i.e. $\Theta_0 = \Theta^{\text{sgn}(\mu)}_0$ and $\lambda_0(\mu) = \text{sgn}(\mu) \mu$.
Since the boundary condition $g(y, \theta)$ is real-valued, it is convenient to separate everything into real and imaginary parts. For compactness, we do not write the explicit $\mu$ and $\theta$ arguments on the $\Omega_k$ and $\Upsilon_k$ (each has arguments $+\theta$ and $|\mu|$). Then,
\begin{align}
f(x, y, \theta) =
     &\int_{0}^{\infty} d \mu \sum_{k \geq 0} \left\{+\Re\left[a_k(\mu)\right]\Omega_k-\Im\left[a_k(\mu)\right]\Upsilon_k+\Re\left[a_k(-\mu)\right]\Omega_k+\Im\left[a_k(-\mu)\right] \Upsilon_k \right\} e^{-\lambda_k x} \cos \mu y \nonumber \\
+  &\int_{0}^{\infty} d \mu \sum_{k \geq 0} \left\{+\Re\left[a_k(\mu)\right]\Upsilon_k+\Im\left[a_k(\mu)\right]\Omega_k+\Re\left[a_k(-\mu)\right]\Upsilon_k-\Im\left[a_k(-\mu)\right] \Omega_k \right\} e^{-\lambda_k x}  \sin \mu y \nonumber \\
+ \, i &\int_{0}^{\infty} d \mu \sum_{k \geq 0} \left\{+\Re\left[a_k(\mu)\right]\Upsilon_k+\Im\left[a_k(\mu)\right]\Omega_k-\Re\left[a_k(-\mu)\right]\Upsilon_k+\Im\left[a_k(-\mu)\right] \Omega_k \right\} e^{-\lambda_k x}  \cos \mu y \nonumber \\
+ \, i &\int_{0}^{\infty} d \mu \sum_{k \geq 0} \left\{-\Re\left[a_k(\mu)\right]\Omega_k+\Im\left[a_k(\mu)\right]\Upsilon_k+\Re\left[a_k(-\mu)\right]\Omega_k+\Im\left[a_k(-\mu)\right]\Upsilon_k \right\} e^{-\lambda_k x}  \sin \mu y \label{temp343323543}
\end{align}
where the $k = 0$ mode is the positive one, i.e. $\Omega_0 = \Omega^{+}_0$ and $\Upsilon_0 = \Upsilon^{+}_0$. By comparison, the first boundary condition can similarly be expressed as a Fourier integral:
\begin{equation}
f(0, y, \theta) = g(y, \theta) = \int_0^{\infty} d\mu \, q(\mu, \theta) \cos \mu y + \int_0^{\infty} d\mu \, p(\mu, \theta) \sin \mu y, \qquad \cos \theta > 0 \label{temp444232100}
\end{equation}
The problem is solved if, for given $\mu >0$, we can choose $a_k(\mu)$ such that the quantities in brackets in Eq. \eqref{temp343323543} match the respective functions $q(\theta, \mu)$ and $p(\theta, \mu)$. This results in four independent equations with four independent sets of coefficients: $\Re\left[a_k(\pm \mu)\right]$ and $\Im\left[a_k(\pm \mu)\right]$. We solve these equations using a least-squares technique which appears to depend on the completeness of the $\Omega_k$ but also requires use of the $\Upsilon_k$. The validity of the procedure is supported by good numerical convergence on a variety of test functions (see Appendix \ref{appendix:2d-problems} and Ref. \cite{WagnerDissertation}.

Here we show results for the cases $g(y, \theta) = 2 + \cos y$ and $g(y, \theta) = e^{-y^2/4}$. The first boundary condition could model ABPs in an infinite channel whose walls have periodically varying absorption energies, whereas the second describes perfectly absorbing walls in the presence of a steady particle source centered around $y=0$. The densities corresponding to these scenarios are shown in Fig. \ref{fig:3d-planar-cos}. As the first boundary condition involves only a single mode of the set of $y$ eigenfunctions, the diffusion solution is easy to write down:
 \begin{equation}
f_d(x, y, \theta) \simeq 1.176 \cdot e^{-x} \left( \Re [ J_0(2i e^{i \theta/2}) ] \cos y + \Im [ J_0(2i e^{i \theta/2}) ] \sin y \right)
\end{equation}
The density is just $\rho_d(x, y) \simeq 3.693 \cdot e^{-x} \,  \cos y$, which compares well with the exact solution (see Fig.~\ref{fig:3d_planar_dtest_gaussian_1}).

\subsubsection{Validation of the diffusion approximation} \label{subsubsec:planar-geometry-b5}

\begin{figure}
  \includegraphics[keepaspectratio=true,scale=0.3]{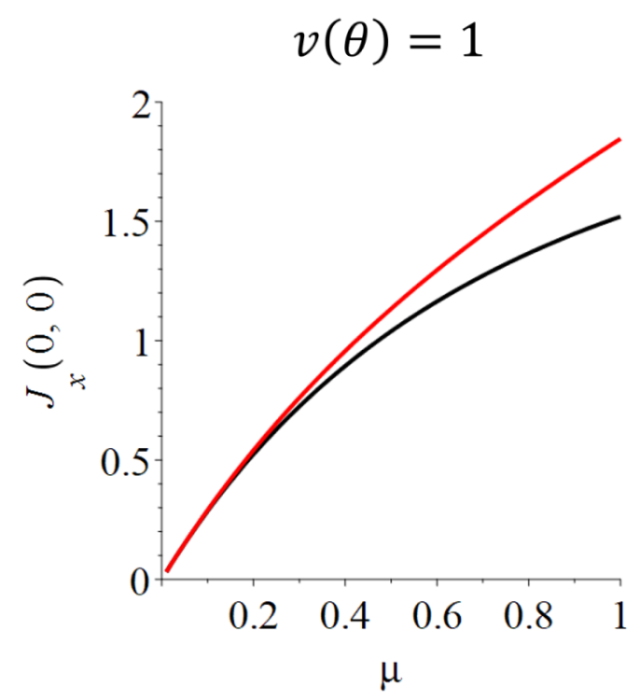} \hspace{2mm}
  \includegraphics[keepaspectratio=true,scale=0.3]{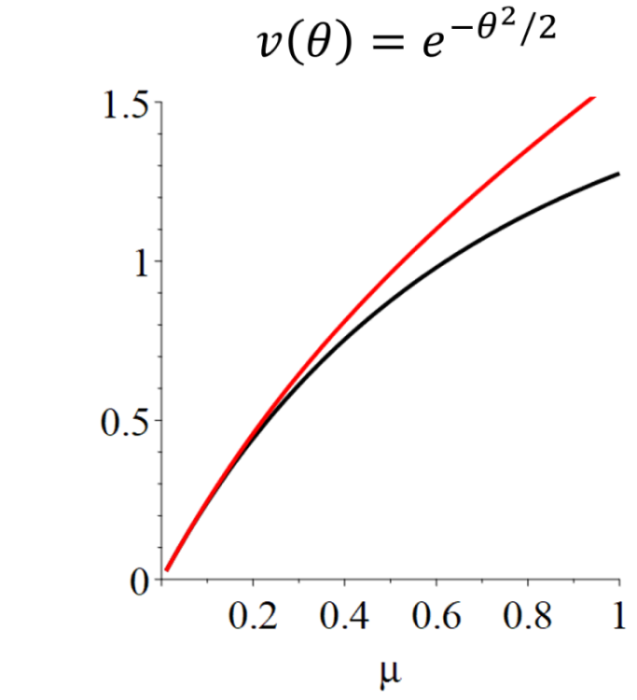} \hspace{2mm}
  \includegraphics[keepaspectratio=true,scale=0.3]{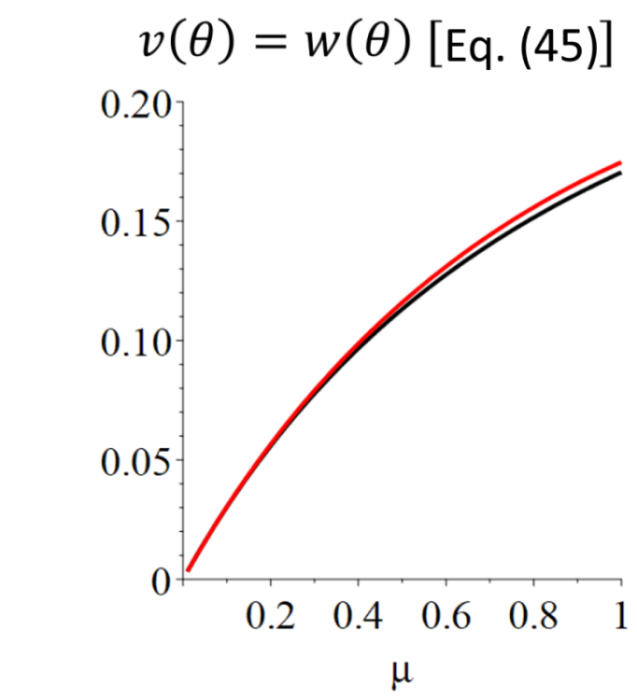} 
  \includegraphics[keepaspectratio=true,scale=0.3]{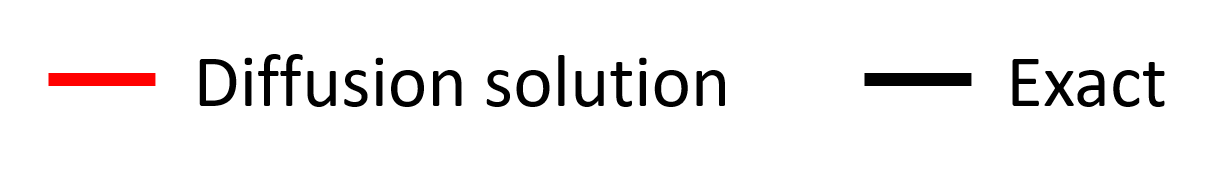}
  \caption{The diffusion approximation says that the diffusion solution carries all the normal flux at the boundary. Here, we test this approximation for the single $\mu$-eigenmode boundary condition $f(0,y,\theta) = v(\theta) \cos \mu y $ in the planar geometry. We find satisfactory agreement for $\mu \ll 1$ ($\mu \ll \ell^{-1}$ in dimensionful form). This supports our claim that the diffusion approximation is valid so long as the length characterizing the variation of the boundary data ($\mu^{-1}$ in this case) is large compared with $\ell$. Since the solution for a general boundary condition $f(0,y,\theta) = g(y, \theta)$ is a sum over single-$\mu$ eigenmodes (Eq. \eqref{eq:temp1543525676}), we expect this conclusion to translate to more general problems as well; see Fig. \ref{fig:3d_planar_dtest_gaussian_1} for examples.}
  \label{fig:3d-planar-dtest-cos}
\end{figure}

\begin{figure}
\centering
  \includegraphics[keepaspectratio=true,scale=0.45]{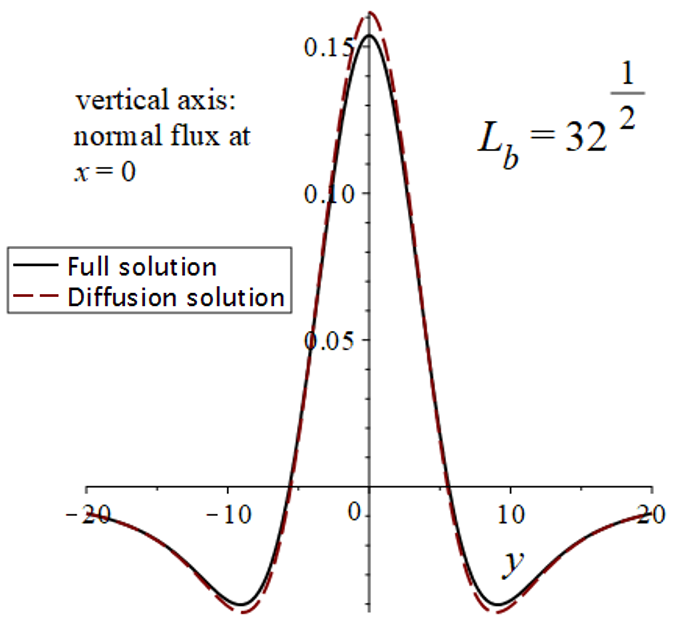} \hspace{5mm} \vspace{15mm}
  \includegraphics[keepaspectratio=true,scale=0.35]{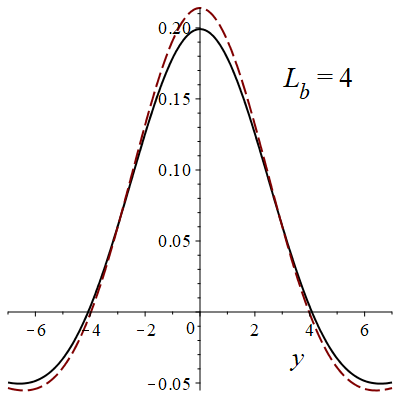}
  \includegraphics[keepaspectratio=true,scale=0.35]{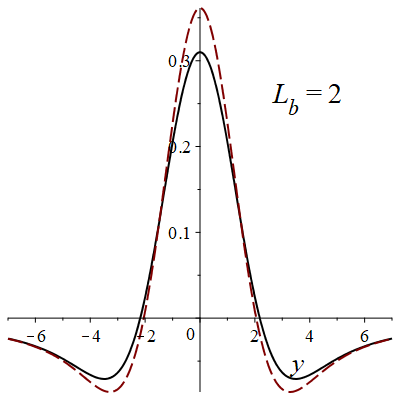} \hspace{5mm}
  \includegraphics[keepaspectratio=true,scale=0.29]{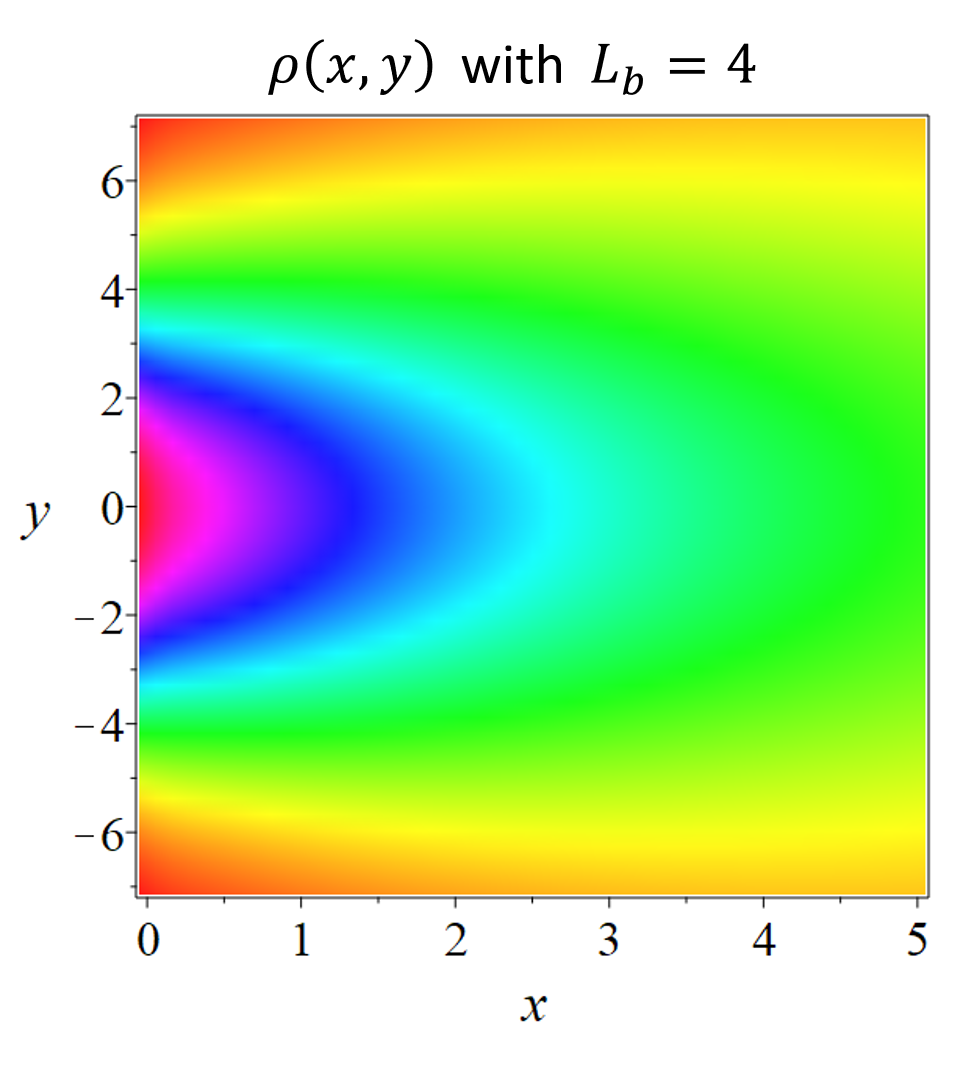} 
    \includegraphics[keepaspectratio=true,scale=0.29]{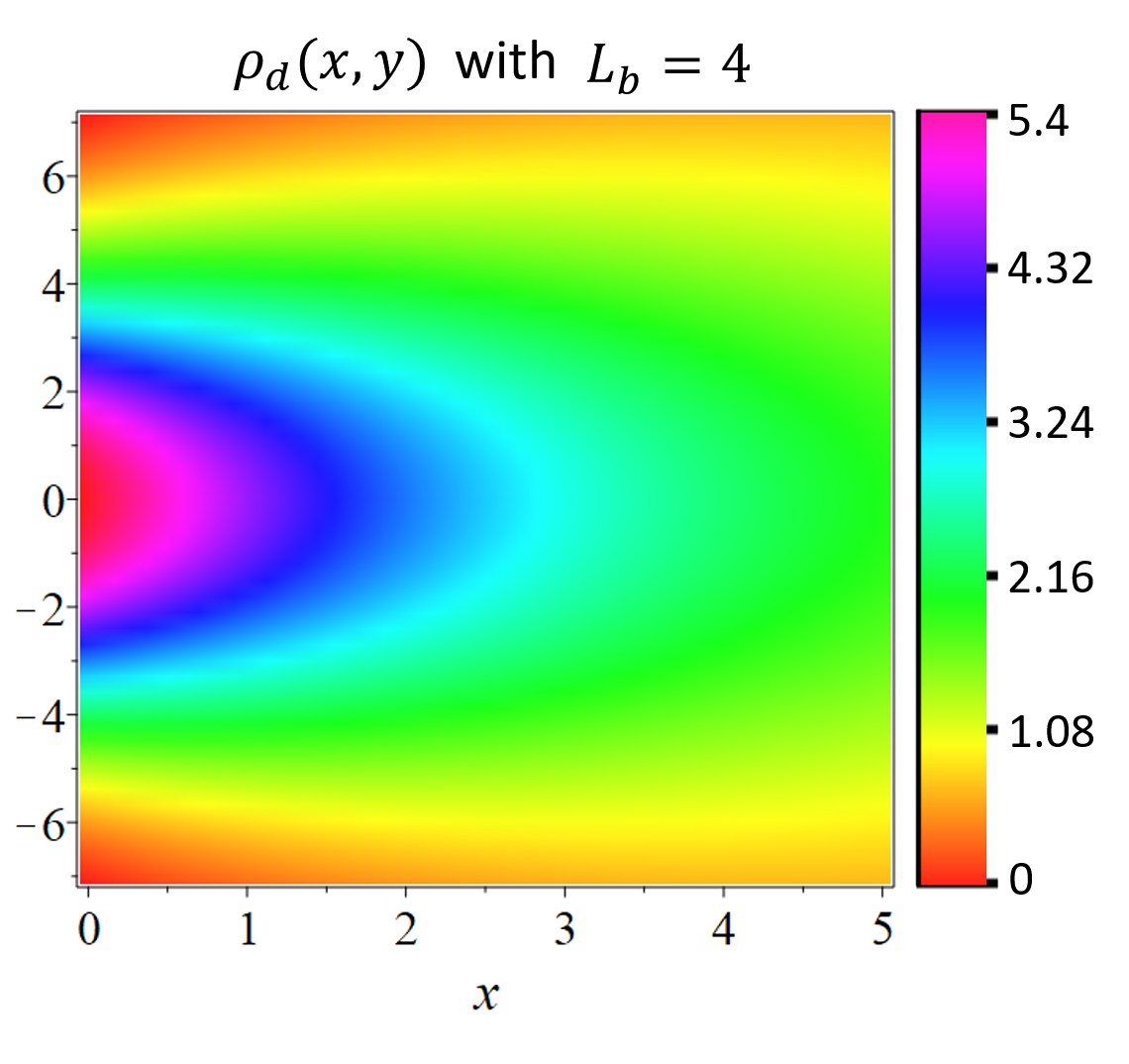}
  \includegraphics[keepaspectratio=true,scale=0.29]{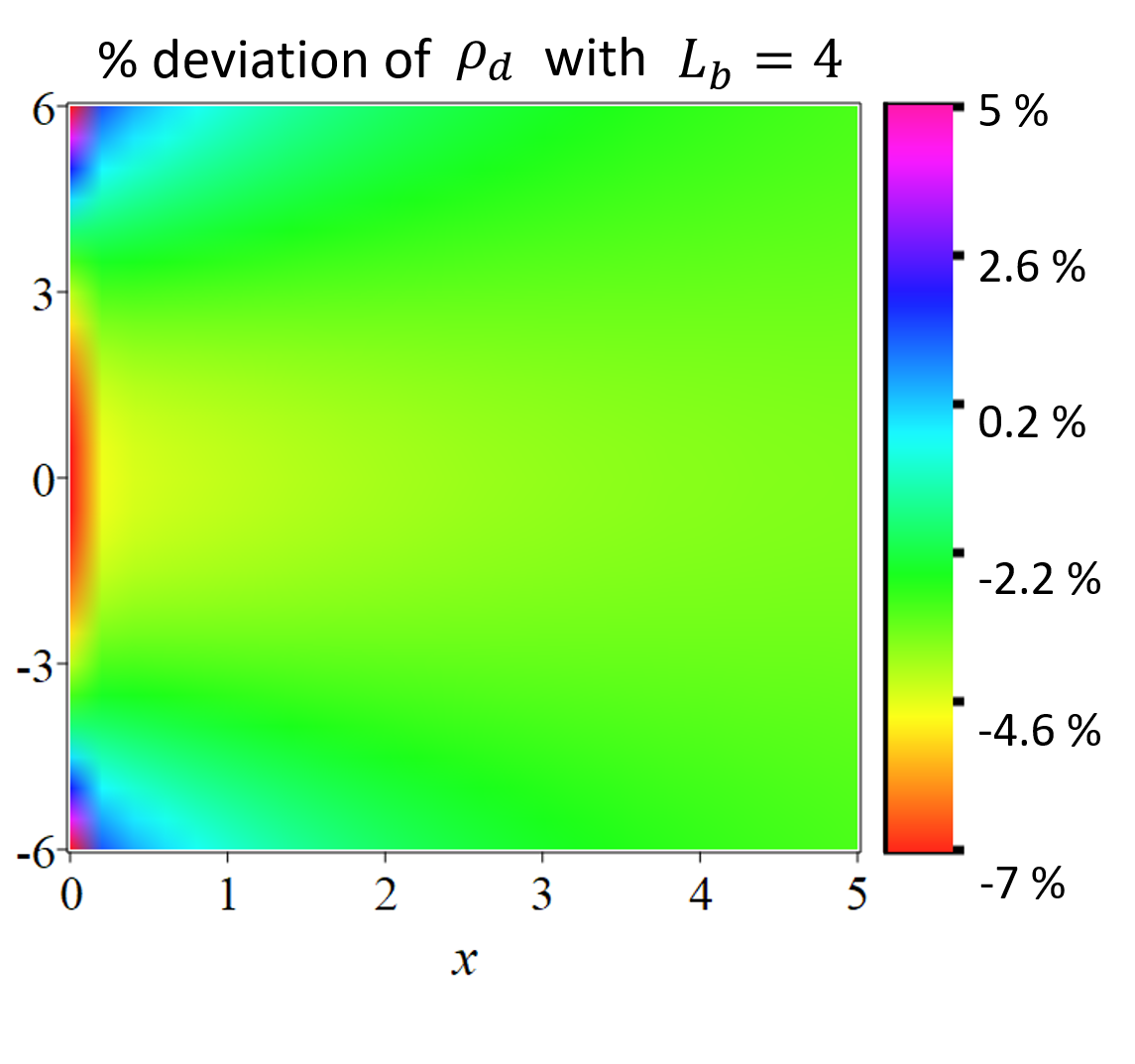} 
  \includegraphics[keepaspectratio=true,scale=0.29]{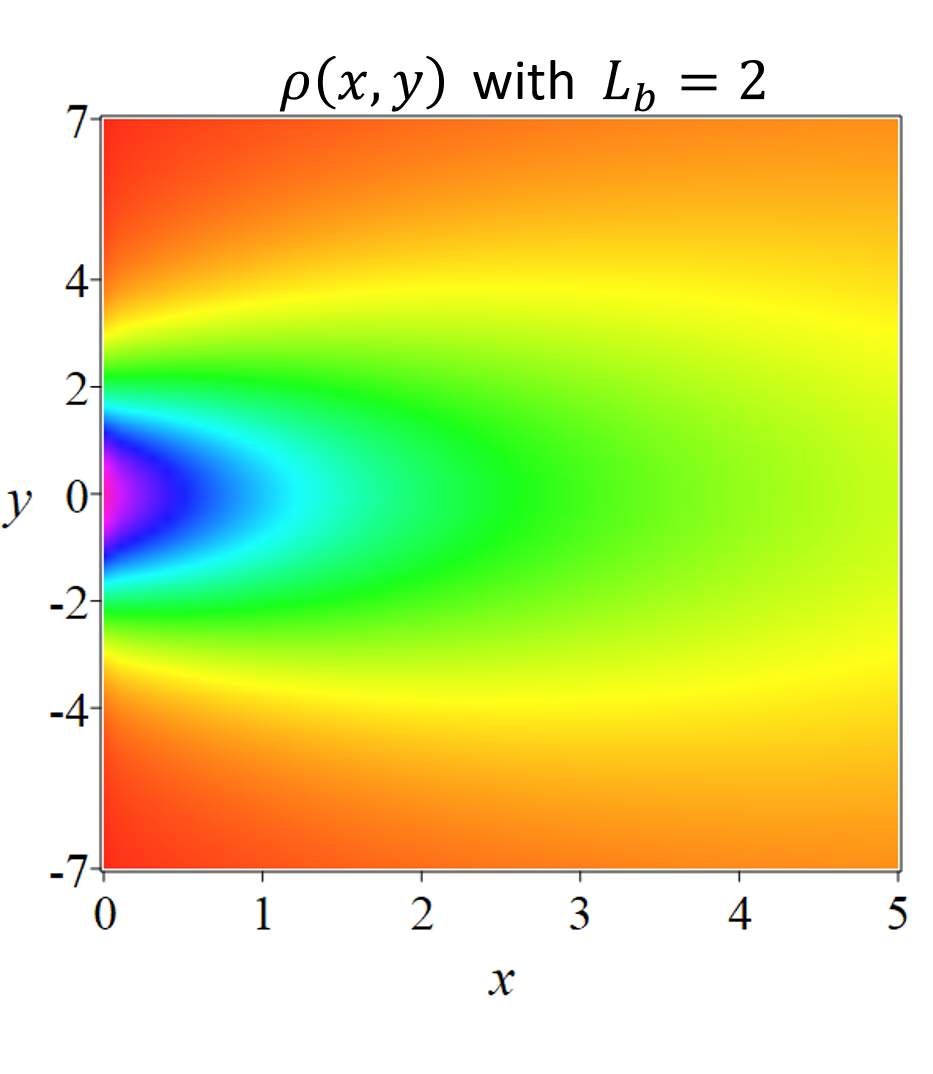}
    \includegraphics[keepaspectratio=true,scale=0.29]{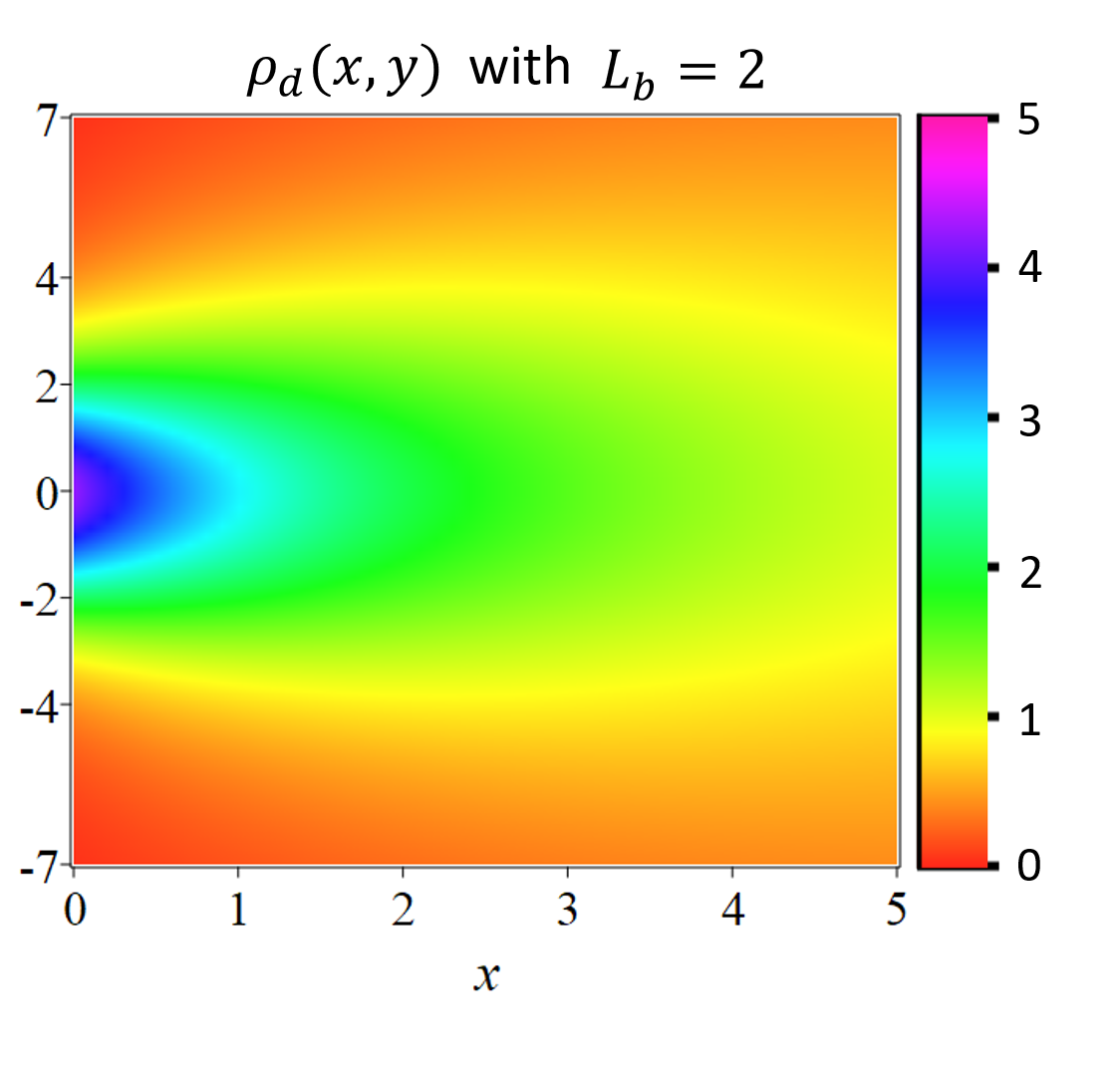}
  \includegraphics[keepaspectratio=true,scale=0.29]{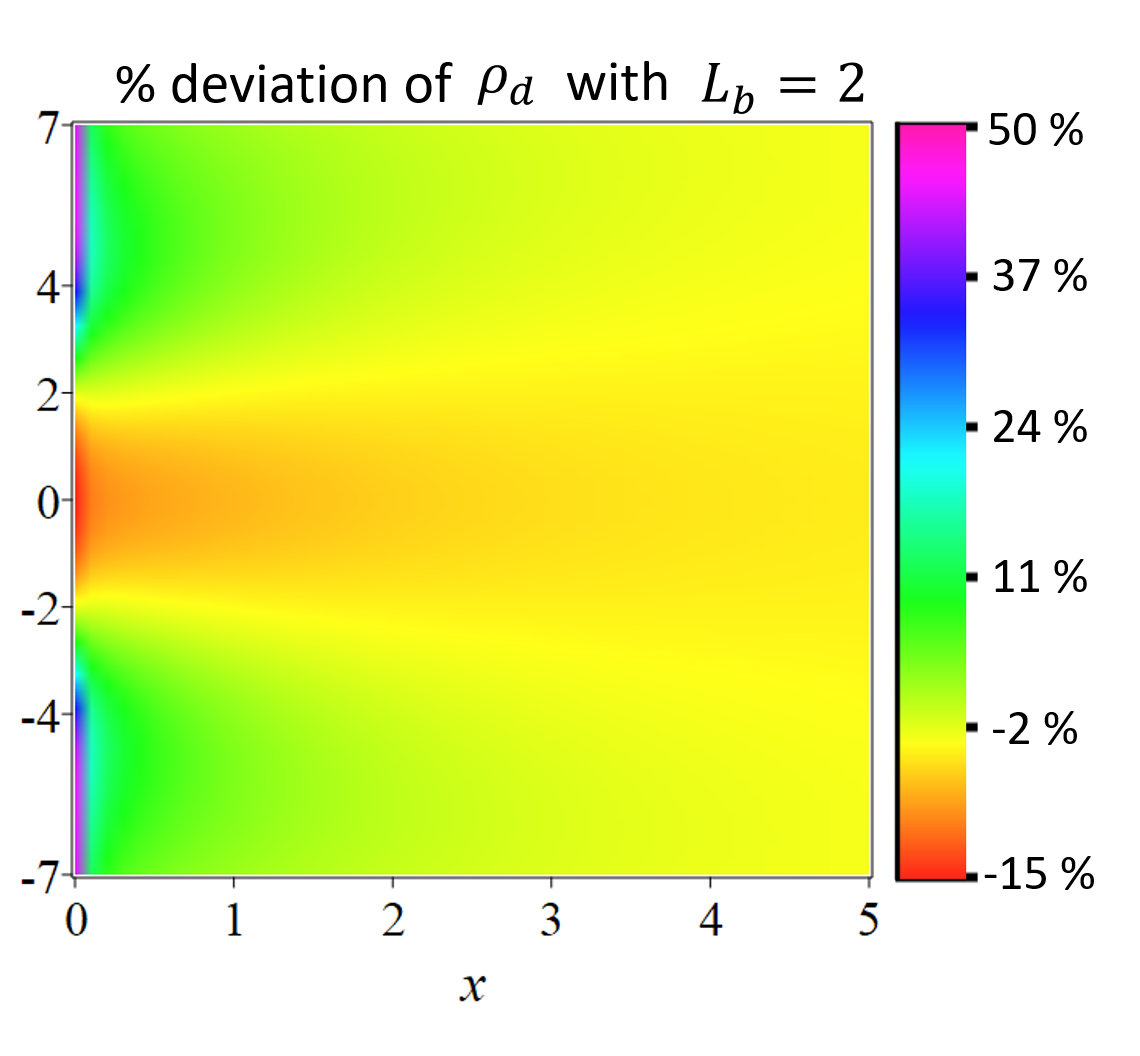}
  \caption[Tests of the diffusion approximation in the planar geometry with boundary condition $f(0,y,\theta) = e^{-y^2/L_b^2}$]{Tests of the diffusion approximation in the planar geometry with boundary condition $f(0,y,\theta) = e^{-y^2/L_b^2}$. The top row shows the normal flux at $x=0$. The diffusion solution indeed carries most of the flux for $L_b > 1$, with the fraction increasing with $L_b$. This supports our claim that the diffusion approximation is valid so long as the length characterizing the variation of the boundary data ($L_b$ in this case) is large compared with $\ell$. In the bottom two rows, we compare the density calculated in the diffusion approximation ($\rho_d$) with the exact density ($\rho$). As expected, the approximation increases in accuracy as $L_b$ becomes large compared with $\ell$.}
  \label{fig:3d_planar_dtest_gaussian_1}
\end{figure}

The validity of the diffusion approximation depends on whether the diffusion solution carries most of the boundary normal flux. The explicit construction of the boundary layer in section \ref{subsubsec:planar-geometry-b3} allows us to test this assumption directly. Working from the representation \eqref{eq:2d-planar-19}, we find
{ \everymath={\displaystyle}
\begin{align}
\int_{-\pi}^{\pi} \Theta_{k}(\theta, \mu) \left[ \begin{array}{ccc} \sin \theta \\ \cos \theta \end{array} \right] d \theta = \left\{
\begin{array}{cc}
\left[ \begin{array}{ccc} 0 \\ 0  \end{array} \right]& \quad k \text{ even (and nonzero)} \vspace{3mm} \\
 \frac{\pi A_{1}^{(k)}(0)}{r_k} \left[ \begin{array}{ccc} \sqrt{|r_k|^2 + \mu^2} \\ -i \mu  \end{array} \right] & \quad  k \text{ odd}
\end{array}
\right. \label{eq:2d-planar-23}
\end{align}
}
Crucially, Eq. \eqref{eq:2d-planar-23} tells us that the boundary normal flux contained in the boundary layer solutions is \emph{at most} $\mathcal{O}(\mu)$, i.e.
\begin{equation}
\int (\text{boundary layer}) (\hat{n} \cdot \hat{u}) d \theta = \int (\text{boundary layer}) \cos \theta \, d \theta = \mathcal{O}(\mu)
\end{equation}
Because $\mu$ is the wavenumber of the $y$-dependent part of the boundary layer (in units of $\ell$), it follows that the assumptions of the diffusion approximation are violated, at most, to order $\mathcal{O}(\ell/L_b)$, where $L_b$ is the length scale describing the variation of the boundary data with $y$.
We have tested this conclusion numerically for a selection of boundary conditions. Notably, with the analytic solution \eqref{temp343323543} in hand, it is possible to analytically separate the diffusion solution from the boundary layer and test the diffusion approximation directly, which is not possible for other geometries considered later. First, we consider the boundary condition $g(y, \theta) = v(\theta) \cos \mu y$ corresponding to a single mode of the set of $y$ eigenfunctions. The solution for general boundary conditions will be a sum over modes of this form, via Eq. \eqref{temp343323543}. Thus, knowing how the accuracy of the diffusion approximation varies with $\mu$ will allow us to judge its accuracy for general problems as well. Towards this end, Fig. \ref{fig:3d-planar-dtest-cos} compares the exact normal flux $J_x$ with the flux of the diffusion solution alone, for three different $v(\theta)$: $v(\theta) = 1$, $e^{-\theta^2/2}$, and
\begin{equation}
v(\theta) =\left( \left( \frac{\pi}{2} \right)^2 - \theta^2 \right) \cdot \left( e^{-5 (\theta - 1.5)^2} +  e^{-5 (\theta + 1.5)^2} \right) \equiv w(\theta)
\end{equation} 
where $w(\theta)$ is chosen to be representative of a smooth function peaked near $\pm \pi / 2$. In all cases, the approximation is good for $\mu \ll 1$, whereas for $\mu > 1$ the accuracy deteriorates.
Thus, the diffusion approximation should work well for boundary conditions which vary slowly with respect to $y$, corresponding to $\mu \ll 1$. Physically, if $L_b$ is the length scale describing the boundary variations, then we require $L_b \gg \ell$. On the other hand, modes with $\mu \gg 1$ decay rapidly away from the boundary, so we might expect the diffusion approximation to fare reasonably even when a small number of modes with $\mu > 1$ are present.

To test these conclusions, we next consider the boundary condition $g(y, \theta) = e^{-y^2/L_b^2}$. Fig. \ref{fig:3d_planar_dtest_gaussian_1} compares the exact normal flux $J_x$ with the flux of the diffusion solution alone for $L_b = 2, 4, 32^{1/2}$. As predicted, the two quantities coincide with high accuracy for $L_b \gg \ell = 1$, and even when $L_b \sim 1$, the agreement is at least qualitative.
Fig. \ref{fig:3d_planar_dtest_gaussian_1} also compares the density from the diffusion approximation with the exact density. As expected, the agreement is poor near $x = 0$, where the boundary layer contribution is significant. Far from the boundary, there are still systematic deviations, but of a lesser degree: around 5 \% relative error for $L_b = 4$ and 10 \% for $L_b = 2$. Moreover, since the deviations are systematic, the qualitative agreement is good. We conclude that the diffusion approximation gives a semi-quantitative picture of the asymptotic density for boundary variations on the order of $\ell$.

\section{Illustration 1: circular geometry} \label{sec:circular-geometry}

\subsection{Problem statement} \label{sec:circular-geometry-a}

We define polar coordinates $(r, \alpha)$ by $x = r \cos \alpha$ and $y = r \sin \alpha$. Applying this transformation to Eq. \eqref{eq:introduction-4} gives
\begin{equation}
\cos (\alpha - \theta) \frac{\partial f}{\partial r} - \frac{\sin(\alpha - \theta)}{r} \frac{\partial f}{\partial \alpha} = \frac{\partial^2 f}{\partial \theta^2}
\label{eq:circular-geometry-1}
\end{equation}
In a geometry which is radially symmetric, $f$ depends on $\alpha$ and $\theta$ only through the combination $\phi \equiv \alpha - \theta$. Then, \eqref{eq:circular-geometry-1} simplifies to
\begin{equation}
\cos \phi \frac{\partial f}{\partial r} - \frac{\sin \phi}{r} \frac{\partial f}{\partial \phi} = \frac{\partial^2 f}{\partial \phi^2}
\label{eq:circular-geometry-2}
\end{equation}
where $-\pi < \phi < \pi$. Perhaps surprisingly, neither of these equations is separable. More generally, we have not found Eq. \eqref{eq:introduction-4} to be separable in any spatial coordinates besides rectangular. Thus, the diffusion approximation is essential for analyzing even simple curvilinear geometries, since the boundary layer can no longer be accessed via a separation of variables.

In this section, we solve the above equations with a variety of boundary conditions. First, we consider an annular region $R_1 < r < R_2$, in which case boundary conditions take the form
\begin{align}
&f(R_1, \alpha, \theta) = B_1(\alpha, \theta) \ \ \ \text{where} \cos \phi > 0 \label{eq:circular-geometry-3} \\
&f(R_2, \alpha, \theta) = B_2(\alpha, \theta) \ \ \  \text{where} \cos \phi < 0 \label{eq:circular-geometry-4}
\end{align}
We consider both spherically symmetric and non-symmetric boundary data, focusing on the qualitative differences between the solutions in either case. Second, we consider the infinite exterior region $r > R_1$, in which case the second boundary condition is replaced by $\lim_{r \rightarrow \infty} f(r, \alpha, \theta) = \text{constant}$.

\subsection{Diffusion solution} \label{sec:circular-geometry-b}

The diffusion solution in the spherically symmetric case of \eqref{eq:circular-geometry-2} is easy to calculate:
\begin{align}
f_d(r, \phi) &= a + b \left(\ln r - \sum_{m=1}^{\infty} \frac{1}{m! \, m} \frac{\cos m \phi}{r^m} \right) \label{eq:circular-geometry-6} \\
&\equiv a + b \, g(r,\phi) \label{eq:circular-geometry-7}
\end{align}
where $a$ and $b$ are constants.
Note that since \eqref{eq:circular-geometry-2} is effectively a 1d problem, the most general diffusion solution contains only two undetermined constants \cite{Kruskal1980,Beals1985}.  In the absence of spherical symmetry, the problem is of course more complicated and in this case, to quadrupole order, we have
\begin{equation}
f_d(r, \alpha, \theta) = c_0 + \beta - \frac{\partial \beta}{\partial r} \cos \phi + \frac{1}{r} \frac{\partial \beta}{\partial \alpha} \sin \phi + \frac{\partial^2 \beta}{\partial r^2} \cos 2 \phi + \frac{1}{r} \frac{\partial^2 \beta}{\partial r \partial \alpha} \sin 2 \phi + \ldots \label{eq:circular-geometry-8}
\end{equation}
where $\phi = \alpha - \theta$ and $\beta(r, \alpha)$ solves $\nabla^2 \beta = 0$. With this in mind, we introduce a multipole expansion for the density $\rho$:
\begin{equation}
\frac{\rho(r, \alpha)}{2 \pi} = c_0 + \beta(r, \alpha) \equiv c_0 - \frac{\boldsymbol{p} \cdot \hat{r}}{r} + \frac{Q_{ij} \hat{r}_i \hat{r}_j}{r^2} + \mathcal{O}\left(\frac{1}{r^3}\right) \label{eq:circular-geometry-9}
\end{equation}
where the constant term has been absorbed into $c_0$, and $\boldsymbol{p}$ and $\mathbf{Q}$ are determined from the boundary conditions on $\beta(r, \alpha)$.

\begin{figure}
  \includegraphics[keepaspectratio=true,scale=0.25]{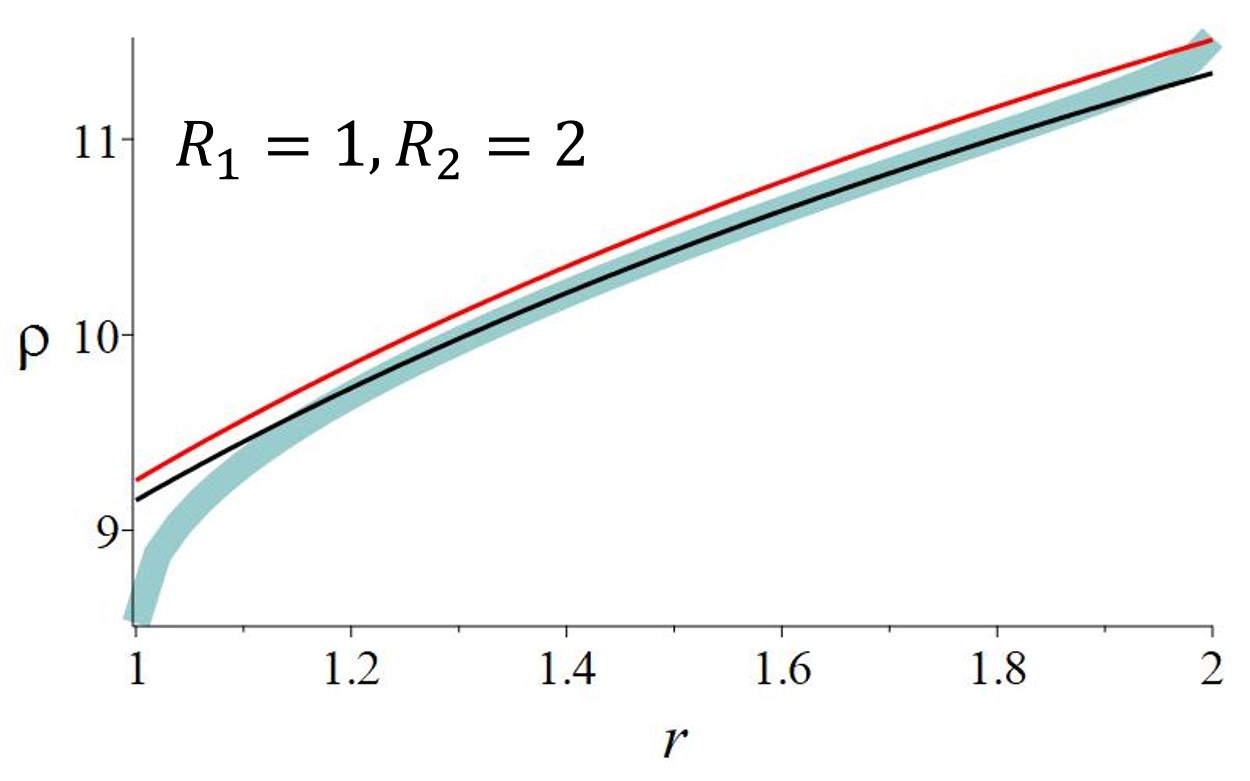} \hspace{5mm}
  \includegraphics[keepaspectratio=true,scale=0.25]{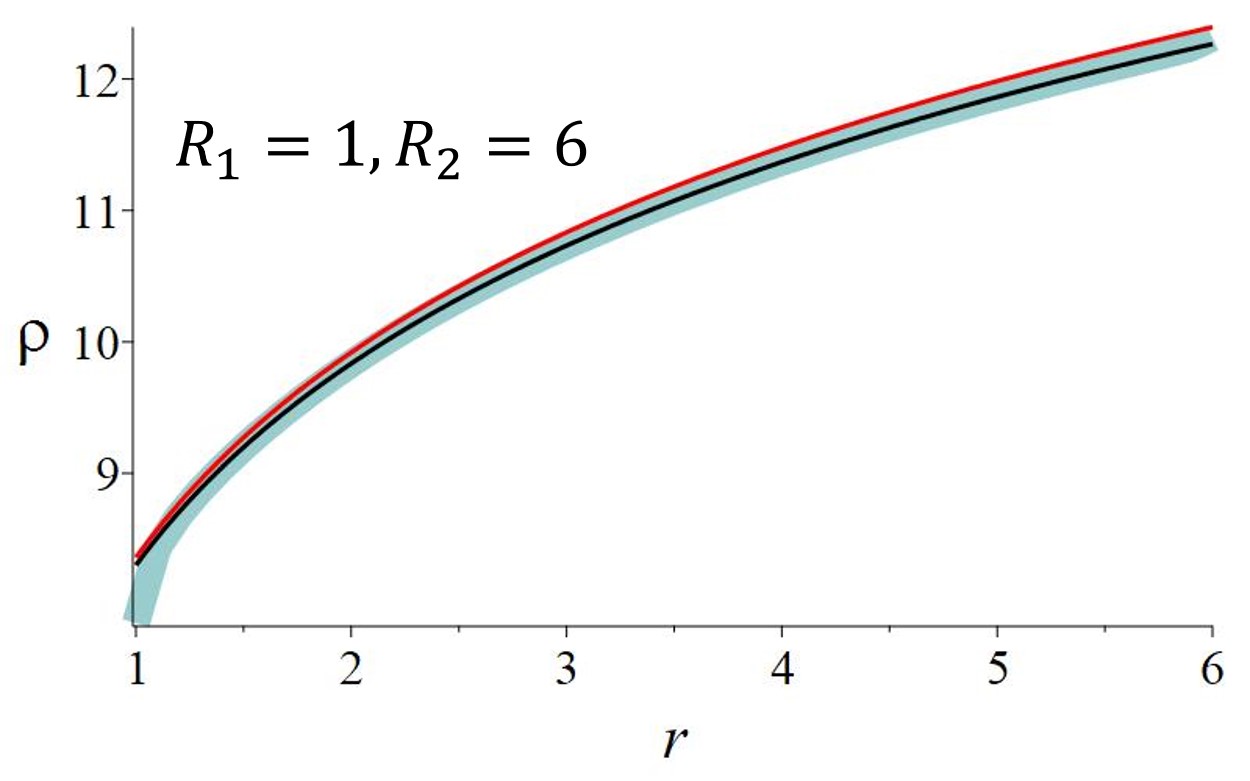} \hspace{5mm}
  \includegraphics[keepaspectratio=true,scale=0.3]{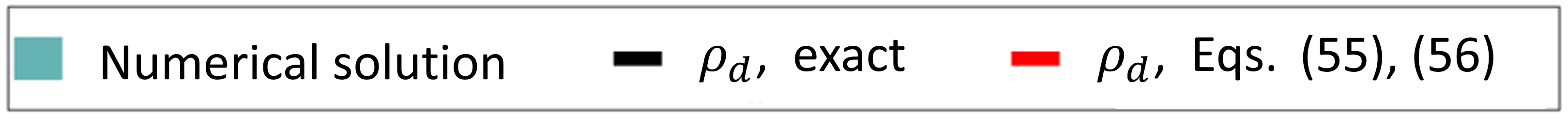}
  \caption{Density profiles for the symmetric annular problem (boundary conditions \eqref{eq:circular-geometry-3}--\eqref{eq:circular-geometry-4} with $B_1 \mathop{=} 1$ and $B_2 \mathop{=} 2$).}
  \label{fig:annular_plots}
\end{figure}

\subsection{Symmetric circular geometry} \label{sec:circular-geometry-c}

In the case of spherical symmetry, the diffusion approximation is exact. The reason can be understood by integrating Eq. \eqref{eq:circular-geometry-2} with respect to $\phi$, which gives
\begin{equation}
\hat{r} \cdot \mathbf{J} = \int f \, \cos \phi \, d \phi = \frac{\mathrm{constant}}{r} \label{eq:circular-geometry-10}
\end{equation}
Comparing with \eqref{eq:circular-geometry-6}, we see that the diffusion solution also has this property. On the other hand, the boundary layer part of the solution decays exponentially with $r$ away from the boundary, and therefore cannot contribute to the overall flux without violating \eqref{eq:circular-geometry-9}. Thus, the diffusion approximation is exact, and the constant $b$ in \eqref{eq:circular-geometry-7} is equal to $ -(\hat{r} \cdot \mathbf{J}) / \pi$. Fig. \ref{fig:annular_plots} compares the density $\rho_d(r)$ obtained in this way with a finite-element solution from PDE2D, verifying its accuracy in the bulk of the region.

An alternate (though approximate) approach estimates the solution directly in terms of the boundary functions $B_1$ and $B_2$, and therefore does not require prior knowledge of $\hat{r} \cdot \mathbf{J}$. The idea is to choose $a$ and $b$ such that the boundary conditions are satisfied ``in the mean'' with respect to weight functions $\cos \phi$ and $\cos^2 \phi$:
\begin{align}
\int_{\cos \phi > 0} \left[ B_1(\phi) - f_d(R_1, \phi) \right] \cos \phi \, d\phi + \int_{\cos \phi < 0} \left[ B_2(\phi) - f_d(R_2, \phi) \right] \cos \phi \, d \phi = 0 \label{eq:circular-geometry-12} \\
\int_{\cos \phi > 0} \left[ B_1(\phi) - f_d(R_1, \phi) \right] \cos^2 \phi \, d\phi + \int_{\cos \phi < 0} \left[ B_2(\phi) - f_d(R_2, \phi) \right] \cos^2 \phi \, d \phi = 0 \label{eq:circular-geometry-13}
\end{align}
As shown in Fig. \ref{fig:annular_plots}, this approximation agrees well with the numerical solution. In fact, in a 1d planar geometry, this approximation is the first term in a convergent iterative scheme \cite{Wagner2019a}, and so we expect it to be generally valid for large $R_1$ and $R_2 \mathop{-} R_1$.

\subsection{Asymmetric circular geometry} \label{sec:circular-geometry-d}

\begin{figure}
  \includegraphics[keepaspectratio=true,scale=0.25]{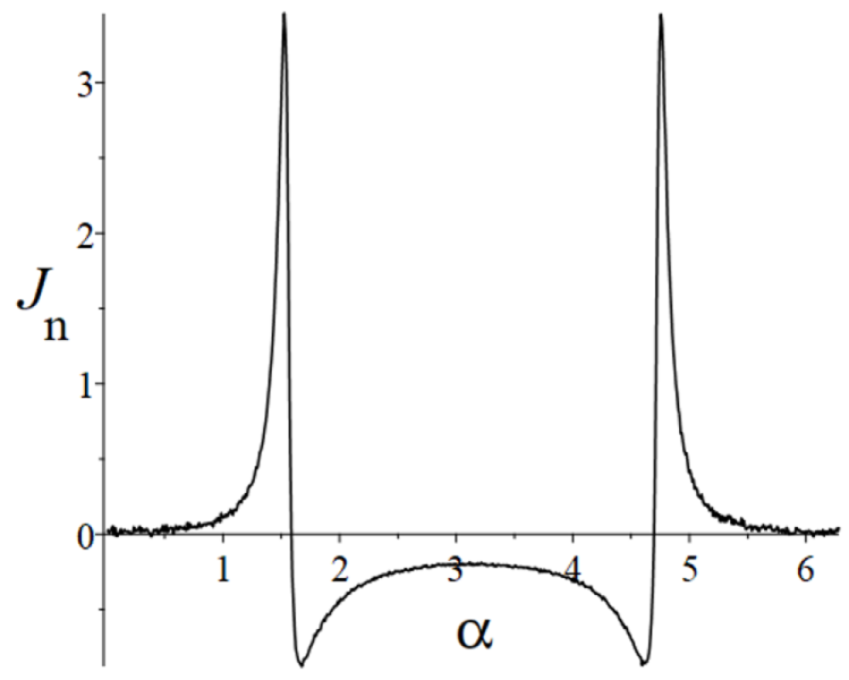}
  \includegraphics[keepaspectratio=true,scale=0.2]{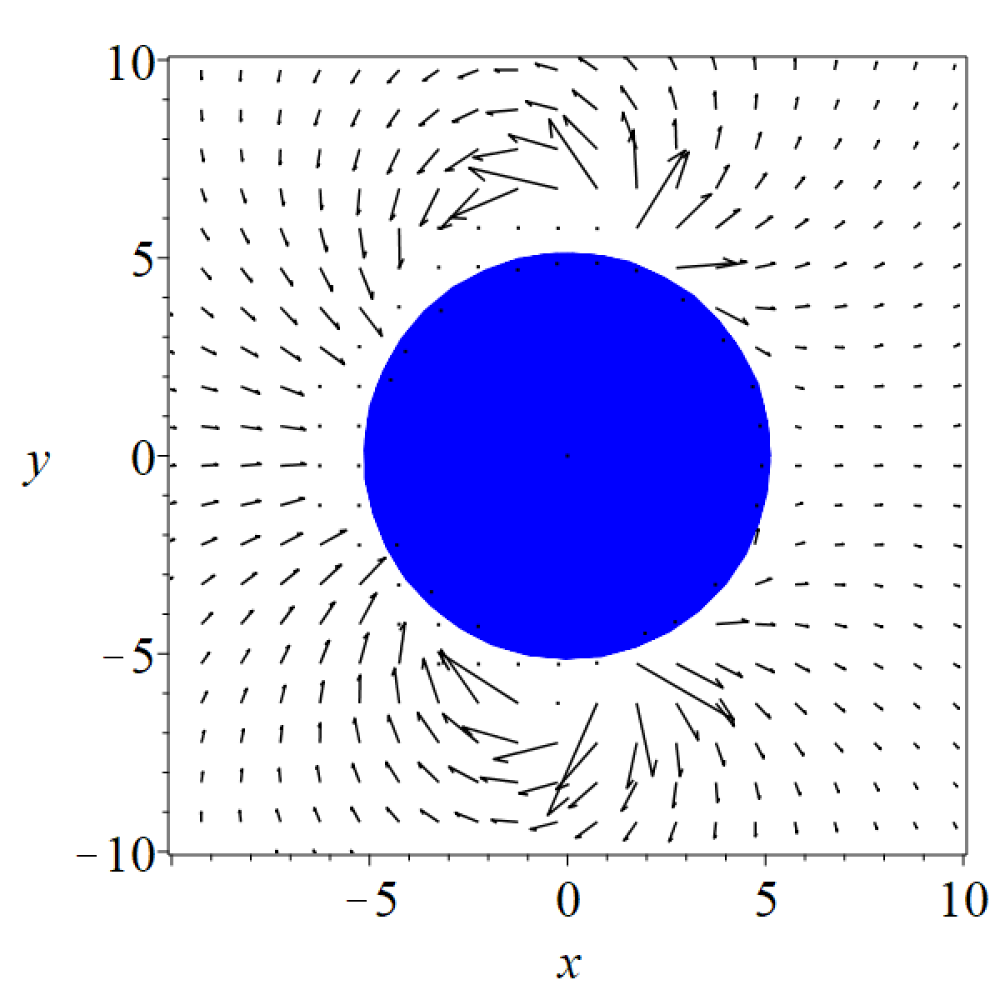}
  \caption{In a closed system of ABPs in steady state, particle-surface interactions that break spherical symmetry induce current loops that begin and end on different points of the boundary. This figure shows an example, in which the particle-surface interactions derive from a continuous, time-independent potential energy function that breaks spherical symmetry (see Eq.~\ref{eq:circular-geometry-14} and appendix A). Qualitatively, this interaction binds particles on the left side of the boundary and releases them on the right side. The left plot shows the normal flux $J_{\text{n}}$ at the boundary (black points) as a function of the angle $\alpha$ with respect to the $x$-axis. On the right is a vector field plot of $J_{\text{n}}$ in the vicinity of the boundary (Flux vectors above a certain threshold magnitude were excluded to improve visualization.)}
  \label{fig:R5_step_particle_flux}
\end{figure}

\begin{figure}
  \includegraphics[keepaspectratio=true,scale=0.2]{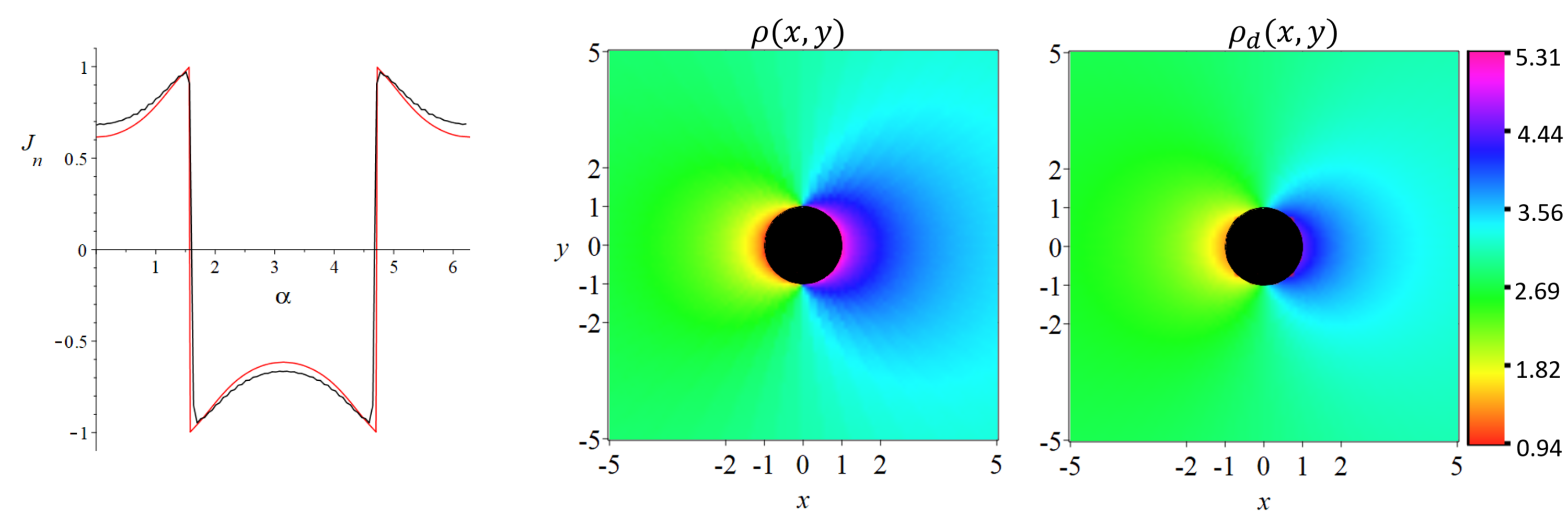}
\caption{Tests of our analytical predictions for boundary conditions \eqref{eq:circular-geometry-14}--\eqref{eq:circular-geometry-15} with $R=1$ and $g_+ = 1$. The left plot compares the boundary normal flux $J_{\text{n}}$ from the finite element solution (black points) with the analytical result (red curve) from the locally planar analysis in appendix \ref{appendix:local-planar}. The other two plots compare the particle density in the diffusion approximation ($\rho_d$, middle plot) with the finite-element solution ($\rho$, left plot). The agreement is quite good given that the analytical approximations have no free parameters.}
  \label{fig:R1_step1}
\end{figure}

In contrast with a spherically symmetric geometry, asymmetric boundary conditions can generate currents and long-range density variations (Fig. \ref{fig:R5_step_particle_flux}). In a system of ABPs, one way to achieve this is with a fixed, spherical inclusion which attracts and binds particles on one side but is purely repulsive on the other side. Then, particles can escape only from one side of the sphere. For concreteness, suppose that the side with binding interactions is the left side, where $x < 0$. Then the boundary conditions are
\begin{align}
f(R, \alpha, \theta) &=
\left\{
\begin{array}{cc}
g_{+}(\alpha, \theta)  & \quad \text{where } \cos \alpha > 0 \text{ and } \cos (\theta -\alpha) > 0 \\
0 & \quad \text{where } \cos \alpha < 0 \text{ and } \cos (\theta -\alpha) > 0 \\
\text{unspecifed} & \text{else}
\end{array}
\right. \label{eq:circular-geometry-14} \\
\lim_{r \rightarrow \infty} f(r, \alpha, \theta) &= c_0 = \text{constant} \label{eq:circular-geometry-15}
\end{align}
such that $g_{+}(\alpha, \theta)$ gives the ``exit distribution" of particles leaving the right side of the sphere. (Since the particles are bound to the left side of the sphere, the corresponding exit distribution there is $0$.) Note that $g_{+}(\alpha, \theta)$ covers only half of the $(\alpha, \theta)$ domain. In fact, on the left side of the sphere, $f(R,\alpha,\theta)$ is nonzero on the complement of the interval where it is $0$ (the ``entrance distribution). However, only the exit distribution in Eq. \eqref{eq:circular-geometry-14} is required to form a well-posed boundary-value problem (see section \ref{sec:model_definitions}).

While the exit distribution is exactly $0$ on the left side of the sphere, the distribution $g_{+}(\alpha, \theta)$ on the right side requires a more detailed analysis that we do not attempt here. For example, one would need to calculate the probability that a trajectory starting on the left side would eventually depart the right side at given $(\alpha,\theta)$. Instead, we test the diffusion approximation independently of the particle-surface kinetics by taking $g_{+}$ to be exactly $1$. In the general case, the diffusion approximation for the density, $\rho_d(r, \alpha)$, is
\begin{align}
\rho_d(r, \alpha) &= 2 \pi c_0 + \sum_{m=1}^{\infty} \left( a_m \frac{\cos m \alpha}{r^m} + b_m \frac{\sin m \alpha}{r^m} \right) \label{eq:circular-geometry-16} \\
\left[ \begin{array}{ccc} a_m \\ b_m \end{array} \right] &= \frac{2}{\pi} \frac{R^{m+1}}{m} \int_{-\pi}^{\pi} J_{\text{n}}(\alpha) \, \left[ \begin{array}{ccc} \cos m \alpha \\ \sin m \alpha\end{array} \right] \, d\alpha \label{eq:circular-geometry-17}
\end{align}
where $J_{\text{n}}(\alpha) \equiv \hat{n} \cdot \mathbf{J}(R, \alpha)$ is the normal flux at the boundary. Ref \cite{WagnerDissertation} works out a strategy for estimating $J_{\text{n}}$ from $g_+$. The idea is to treat the boundary as locally planar and use the machinery from section \ref{subsubsec:planar-geometry-b3} to compute the solution close to the boundary (see Appendix B). Fig. \ref{fig:R1_step1} compares $J_{\text{n}}(\alpha)$ calculated in this way for $R=1$ with the numerical solution from PDE2D. The agreement is surprisingly good considering that $ell$ is comparable to $R$. For larger $R$, the approximation is even better \cite{WagnerDissertation}. In particular, we find the simple expression $J_{\text{n}}(\alpha) \approx (R \cos \alpha)^{-1}$ to be accurate for $R \gg 1$, which in the multipole expansion for the density, Eq. \eqref{eq:circular-geometry-9}, implies a dipole moment
\begin{equation}
\mathbf{p} = (a_1, b_1) = (4R, \, 0), \qquad (R \gg 1). \label{eq:circular-geometry-18}
\end{equation}
Fig. \ref{fig:R1_step1} compares the resulting $\rho_d$ with the finite-element solution. If we exclude the region within a persistence length of the boundary (white annulus), the maximum percent deviation of $\rho_d$ is about $5 \%$. Thus, the diffusion approximation works well in a nontrivial curvilinear geometry, where an exact solution by separation of variables is no longer possible.

\begin{figure}
\floatbox[{\capbeside\thisfloatsetup{capbesideposition={left,top},capbesidewidth=8cm}}]{figure}[\FBwidth]
{\caption{Elliptic geometry. We define elliptic coordinates $(\mu, \eta)$ by $x \mathop{=} c \sinh \mu \cos \eta$ and $y \mathop{=} c \cosh \mu \sin \eta$, where $c \mathop{=} \sqrt{b^2 - a^2}$. These are orthogonal coordinates, and the constant $\mu$ surfaces are confocal ellipses. In particular, $\mu \mathop{=} \mu_0 \equiv \text{arccosh}(b/c)$ defines the ellipse depicted on the right. The blue arrows are the unit tangent and normal vectors at the point $(\mu_0, \eta)$; the unit normal vector is the same as the coordinate vector $\hat{e}_{\mu}$, and similarly the tangent vector is $\hat{e}_{\eta}$. $Z^{-1} = (\cosh^2 \mu_0 \cos^2 \eta + \sinh^2 \mu_0 \sin^2 \eta)^{-1/2}$ is a normalization factor.}\label{fig:test}}
  {\includegraphics[keepaspectratio=true,scale=0.4]{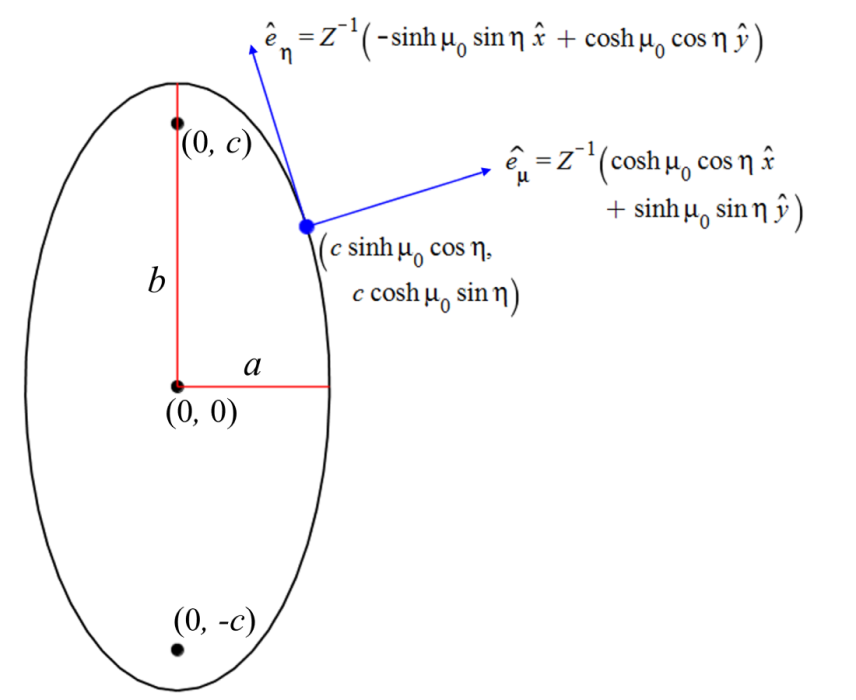}
  \label{fig:ellipse_geometry}}
\end{figure}

\section{Illustration 2: elliptic geometry} \label{sec:elliptic-geometry}

In the polar geometry, an asymmetry in the particle-surface interaction is required to generate long-range flow; otherwise, the net particle current is everywhere $0$ and density decays exponentially away from the boundary. In particular, a purely repulsive spherical inclusion does not generate currents. In this section, we turn to an elliptic geometry, in which the requisite asymmetry is present in the shape of the boundary itself. Experimentally, such geometries would allow one to engineer long-range flows without having to tune particle-surface interactions.

\subsection{Problem setup}

Fig. \ref{fig:ellipse_geometry} shows the setup: we take our boundary to be a vertically oriented ellipse $E$ with semimajor axis $a$, semiminor axis $b$, and foci at $(0, \pm c)$ where $c \equiv \sqrt{b^2 - a^2}$. We define elliptic coordinates $(\mu, \eta)$,
\begin{equation}
\left\{
\begin{array}{cc}
x = c \sinh \mu \cos \eta & \qquad  0 \leq \mu < \infty \text{ and } 0 \leq \eta < 2 \pi  \\
y = c \cosh \mu \sin \eta
\end{array}
\right.  \label{eq:elliptic-geometry-1}
\end{equation}
such that $E$ is defined by the constant-$\mu$ surface $\mu = \mu_0 \equiv \text{arccosh} (b / c)$. The diffusion solution can be written in terms of $\mu$ and $\eta$ using the general expression
\begin{equation}
f_d(\mathbf{r}, \theta) = c_0 + \beta(\mathbf{r}) - \epsilon^2 \, \boldsymbol{\nabla} \beta(\mathbf{r}) \cdot \hat{u} + \frac{\epsilon^3}{4} \left( \boldsymbol{\nabla} \boldsymbol{\nabla} \beta(\mathbf{r}) \right)_{ij} \hat{u}_i \hat{u}_j + \ldots \label{eq:elliptic-geometry-2}
\end{equation}
where $\nabla^2 \beta(\mathbf{r}) = 0$. Conveniently, the Laplacian is just $\nabla^2 = \partial_{\mu}^2 + \partial_{\nu}^2$, so that $\beta(\mu, \eta)$ can be expanded as
\begin{equation}
\beta(\mu, \eta) = \sum_{m=1}^{\infty} e^{-m \mu} \left(a_m \cos m \eta + b_m \sin m \eta \right) \label{eq:elliptic-geometry-5}
\end{equation}

In the diffusion approximation, the boundary condition is
\begin{equation}
\left. (\hat{e}_{\mu} \cdot \boldsymbol{\nabla} \beta (\mu, \eta)) \right|_{\mu = \mu_0} = -J_{\text{n}}(\eta) \quad \rightarrow \quad \frac{1}{c \left( \sinh^2 \mu_0 + \cos^2 \eta \right)^{1/2}} \frac{\partial \beta}{\partial \mu} = -J_{\text{n}}(\eta) \label{eq:elliptic-geometry-6}
\end{equation}
where $\hat{e}_{\mu}$ is the $\mu$-coordinate vector (which is the same as the unit normal vector) and $J_{\text{n}}(\eta)$ is the normal flux. Thus,
\begin{align}
a_m = \frac{c}{m \pi} e^{m \mu_0} \int_{0}^{2 \pi} \cos m \eta \left(\sinh^2 \mu_0 + \cos^2 \eta \right)^{1/2} J_{\text{n}}(\eta) \, d \eta \label{eq:elliptic-geometry-8}
\end{align}
and similarly for $b_m$. Transforming from $(\mu, \eta)$ to $(x,y)$ gives
\begin{equation}
\beta(\mu, \eta) = \frac{\boldsymbol{p} \cdot \hat{r}}{r} + \frac{Q_{ij} \hat{r}_i \hat{r}_j - (1/2) \mathrm{Tr} \, \mathbf{Q}}{r^2} + \mathcal{O}\left(\frac{1}{r^3}\right) \label{eq:elliptic-geometry-12}
\end{equation}
where
\begin{align}
\boldsymbol{p} = \frac{c}{2} \left(a_1, \, b_1 \right); \qquad \mathbf{Q} = \frac{c^2}{4}\left(
\begin{array}{cc}
a_2 & b_2 \\
b_2 & a_2
\end{array}
\right)  \label{eq:elliptic-geometry-13}
\end{align}

\begin{figure}
  \includegraphics[keepaspectratio=true,scale=0.3]{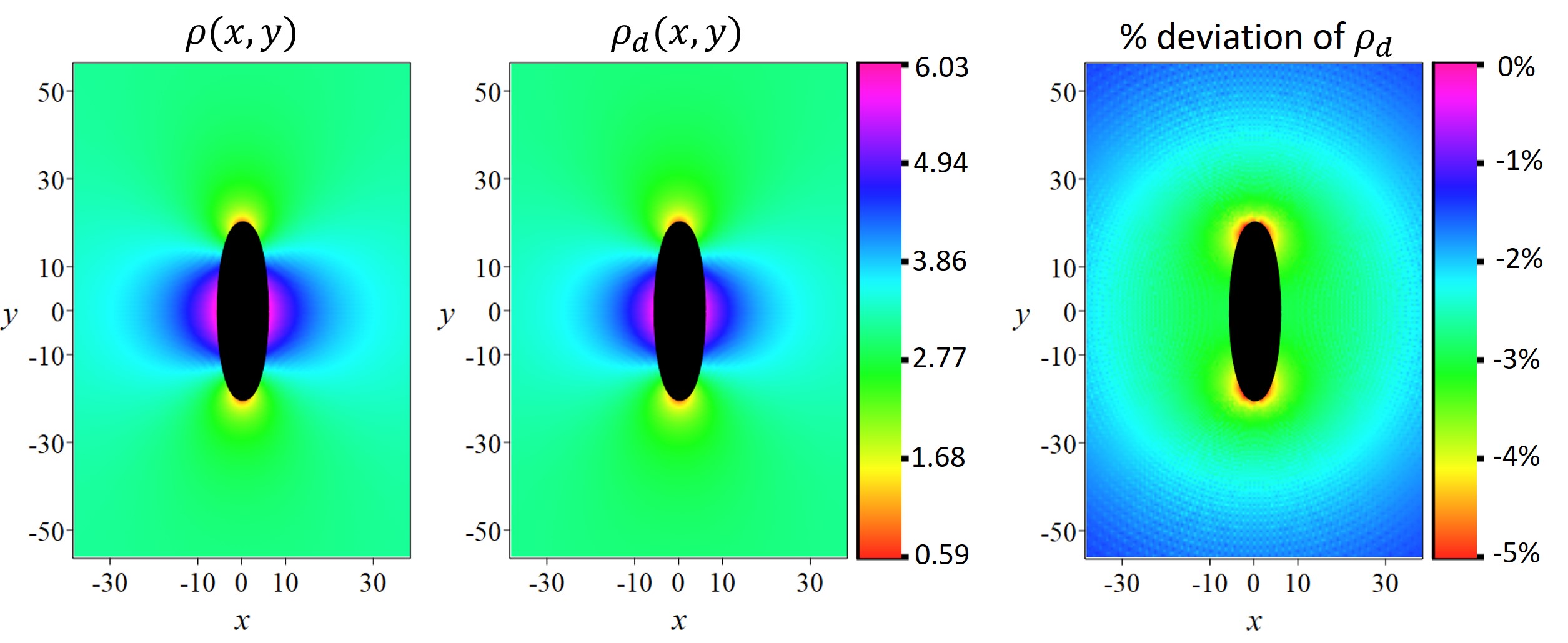}
  \caption{Density profiles corresponding to the boundary condition $f(\mu_0, \eta, \theta) = \cos^2 \eta$ and $f \rightarrow \text{constant}$ as $\mu \rightarrow \infty$. The $\mu = \mu_0$ coordinate surface defines an ellipse (black) with semimajor axis $b=20 \ell$ and semiminor axis $a =5 \ell$. The left and middle plots show the density from PDE2D and the diffusion approximation, respectively. The rightmost plot shows the percent deviation, which is systematic but small.}
  \label{fig:ellipse_pde2d_compare}
\end{figure}

\subsection{Test of the diffusion approximation}

\begin{figure}
  \includegraphics[keepaspectratio=true,scale=0.32]{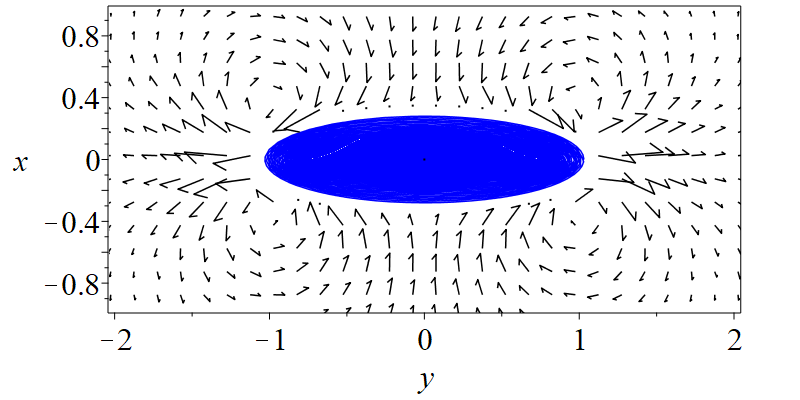} \hspace{3mm} \vspace{4mm}
  \includegraphics[keepaspectratio=true,scale=0.32]{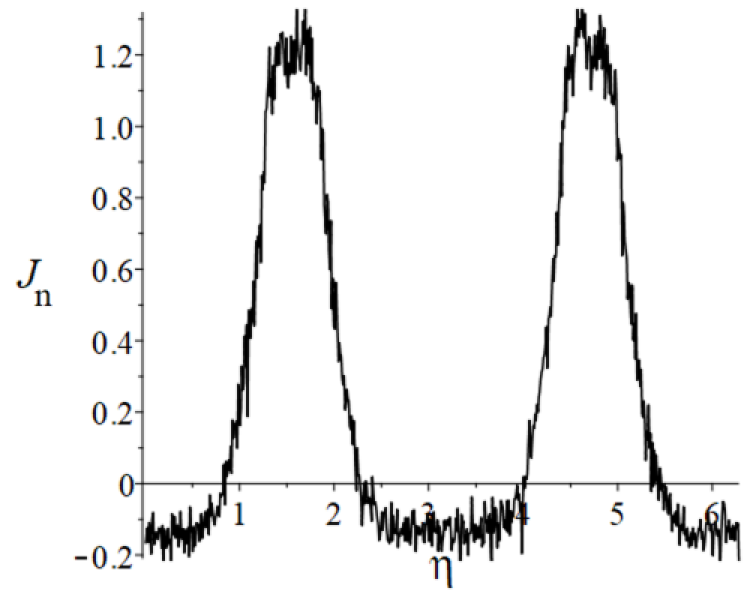}  \vspace{4mm}
  \includegraphics[keepaspectratio=true,scale=0.29]{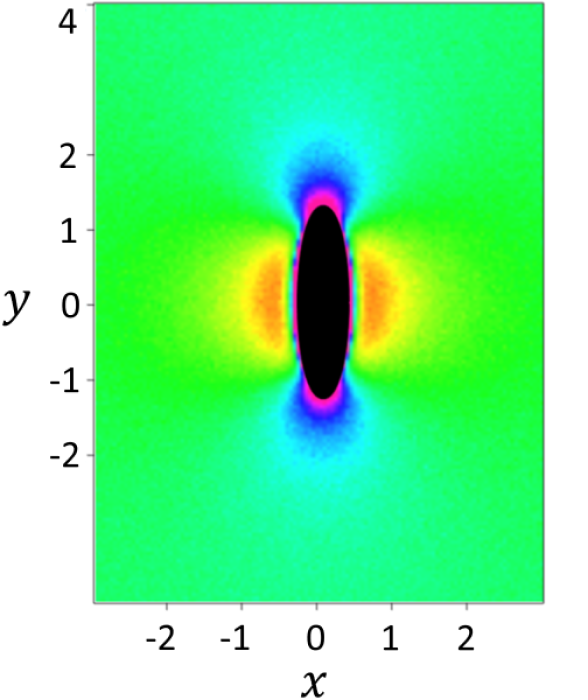}
  \includegraphics[keepaspectratio=true,scale=0.29]{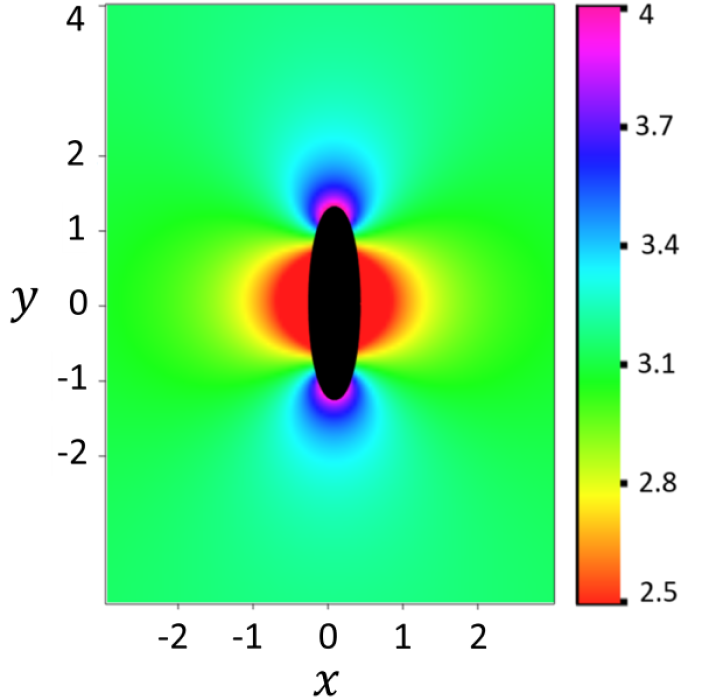}
  \includegraphics[keepaspectratio=true,scale=0.29]{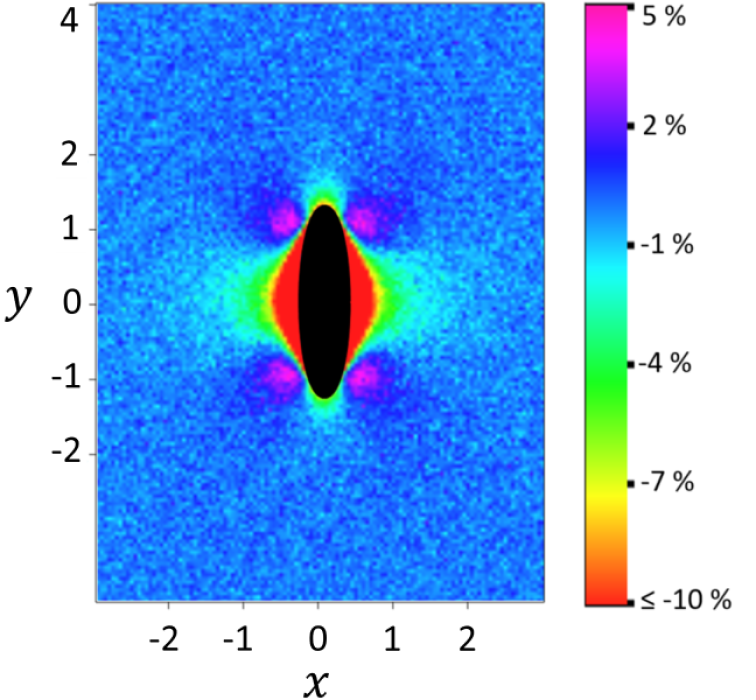}
  \caption{Flux field and density for a bath of ABPs in contact with a fixed, purely repulsive ellipse with semimajor axis $b = \ell$ and semiminor axis $a = 0.25 \ell$. The radius of an ABP is $0.1 \ell$. The top two plots show the flux field and normal flux $J_{\text{n}}$ obtained from simulation of the particle dynamics \eqref{eq:introduction-1}-\eqref{eq:introduction-2}. The bottom left plot shows the density field from simulations. The bottom middle plot shows the diffusion approximation using the boundary condition \eqref{eq:elliptic-geometry-16}. The agreement is fairly good considering the simplifications leading up to \eqref{eq:elliptic-geometry-16}.}
  \label{fig:ellipse_particle_compare}
\end{figure}

Fig. \ref{fig:ellipse_pde2d_compare} compares the diffusion solution with PDE2D for $a = 10$, $b=50$, and the boundary condition
\begin{align}
f(\mu_0, \eta, \theta) &= \cos^2 \eta \label{eq:elliptic-geometry-14} \\
\lim_{\mu \rightarrow \infty} f(\mu, \eta, \theta) &= \text{constant} \label{eq:elliptic-geometry-15}
\end{align}
The agreement is on par with what we saw for the circular geometry. Of course, for a physical system the boundary conditions must be derived self-consistently from the underlying equations of motion, \eqref{eq:introduction-1}-\eqref{eq:introduction-2}. To apply the diffusion approximation, in particular, we need to know the boundary normal flux. Here we present a kinetic argument which approximates this quantity in the case of a long and thin ellipse ($a / b$ small), and moderate to large persistence length $\ell$.

The idea is that in these circumstances, a particle which adsorbs onto the surface tends to drive itself toward the ends of the ellipse before it significantly changes its orientation. This implies that most particles leave the surface in the vicinity of the points $(0, \pm b)$. By contrast, if $\ell$ were much smaller than $b$, then particles would have sufficient time to reorient themselves and leave before they travel too far from where they got adsorbed. To know the net flux, we also have to estimate the incoming flux of particles. Here we simply take this to be constant along the surface: the idea is that a particle returning to the surface has spent enough time in the bulk to forget where it left the surface. Again, we expect this to be a reasonable approximation for sufficiently large $\ell$.

From the above, we are led to the following expression for the boundary normal flux:
\begin{align}
J_{\text{n}}(\eta) &= \frac{\chi}{2a} \left[ \delta \left(\eta - \frac{\pi}{2} \right) + \delta \left(\eta - \frac{3 \pi}{2} \right) - \frac{2a}{c \, I_0} \right] \label{eq:elliptic-geometry-16} \\
\text{where  } I_0 &\equiv \int_{0}^{2 \pi} \left(\sinh^2 \mu_0 + \cos^2 \eta \right)^{1/2} \, d \eta \nonumber
\end{align}
The $\delta$-functions model the outgoing flux due to particles leaving near $(0, \pm b)$, and the constant offset $(1/\pi)$ accounts for the incoming flux, its value chosen so that the net flux integrated along the surface is $0$. The function $F(\eta)$ expresses these two contributions up to an undetermined constant $\chi$, which has units of inverse time. Physically, $\chi$ quantifies the number of particles impinging on (or leaving) the surface in unit time. With this interpretation, a rough guess is that $\chi \approx 2 c_0 \mathcal{A} \, v_0$, $v_0$ being the self-propulsion velocity, $c_0$ the (constant) value of $f$ at infinity, and $\mathcal{A}$ the surface area.

Fig. \ref{fig:ellipse_particle_compare} compares this prediction with simulation for $a = 0.25 \ell$, $b = \ell$, and particle radius $0.1 \ell$. The measured normal flux reflects the general shape of $F(\eta)$, showing peaks near $\cos \eta = 0$ and a constant offset. The predicted density $\rho_d$ using Eq. \eqref{eq:elliptic-geometry-16} also compares well with simulation: most of the error is concentrated in the near field, particularly around $(\pm a, 0)$. In the multipole expansion for the density (Eq. \eqref{eq:elliptic-geometry-12}), the predicted dipole moment is $0$, and the quadrupole moment is $Q_{12} = Q_{21} = 0$,
\begin{align}
Q_{11} = Q_{22} &= -\frac{\mathcal{A}}{8 \pi} \left( a+b \right)^2 \left( 1 + \frac{I_2}{I_0} \right) \label{eq:elliptic-geometry-17} \\
I_m &\equiv \int_{0}^{2 \pi} \cos m \eta \left(\sinh^2 \mu_0 + \cos^2 \eta \right)^{1/2} \, d \eta \label{eq:elliptic-geometry-18}
\end{align}
In the limit $b \gg a$, this becomes $Q_{11} = Q_{22} \approx  (2 \pi)^{-1}\left( a+b \right)^2 \mathcal{A} $. For the problem in Fig. \ref{fig:ellipse_particle_compare}, the numerical value is $Q_{11} \simeq -0.42$.

A caveat is that the predicted flux, Eq. \eqref{eq:elliptic-geometry-16}, is larger than the measured flux by roughly an order of magnitude. Despite this discrepancy, using \eqref{eq:elliptic-geometry-16} as the boundary condition gives a more accurate prediction than using the measured flux. A likely explanation is that the fraction of the flux contained in the exact diffusion solution is greater than $1$, the excess being balanced by the boundary layer piece of the solution. Then, the better agreement of \eqref{eq:elliptic-geometry-16} is due to a cancellation of two errors: the first overestimating the flux and the second underestimating the fraction of flux contained in the diffusion solution. Such situations are expected when $\ell$ is comparable to the length scale of the boundary conditions. Nevertheless, our results suggest that the diffusion approximation may give a qualitative understanding of the steady state even in these cases.

\section{Summary} \label{sec:summary}

Our goal has been to develop an analytical framework to understand ABP steady states in the presence of boundaries. A key simplification that enables this effort is the diffusion approximation, Eq. \eqref{eq:sec3-13}, which says that the asymptotic (``diffusion") solution of \eqref{eq:introduction-4} carries all the particle flux. Our results from sections \ref{sec:planar-geometry}--\ref{sec:elliptic-geometry} demonstrate the validity of this approximation when variations of the boundary data are small compared with the persistence length $\ell$ (set to $1$ in the majority of our analysis). More generally, the diffusion approximation provides a map between violations of detailed balance at the boundary and corresponding mass fluxes in the bulk. This mapping could be a useful design principle for nonequilibrium steady states: by tuning particle-boundary interactions or boundary geometry, one could achieve desired properties of the bulk steady state.

The extension of our framework to higher-dimensional problems could be especially beneficial, as a way to mitigate the accompanying ``curse of dimensionality". Even in cases where higher accuracy is desired than provided by the diffusion approximation, the general form of the asymptotic solution could be used to accelerate numerical algorithms \cite{Jin1999}. Along these lines, a study of long-range flow fields in 3d ABP systems would be interesting.

Similarly, the extension to time-dependent problems would be interesting. The characteristics of the long-time approach to diffusion have lately been considered in the active matter context \cite{Sevilla2015,Malakar2018,Scholz2018,Basu2019,VillaTorrealba2020}, and an asymptotic approach such as ours could provide additional insight. For the ABPs, a possible starting point is to scale Eq. \eqref{eq:introduction-4} as
\begin{equation}
\epsilon^2 \partial_t f + \epsilon \left (\cos \theta \, \partial_x f + \sin \theta \, \partial_y f \right) = \partial_{\theta}^2 f
\end{equation}
where $\epsilon$ is a formal perturbation parameter. Then, substituting $f \mathop{\sim} f_0 + \epsilon f_1 + \ldots \,$ gives $f \mathop{=} \text{const.} \mathop{+} \beta(x, y, t) \mathop{+} \mathcal{O}(\epsilon)$, where $\beta$ solves $\partial_t \beta \mathop{=} (1/4) \nabla^2 \beta$. Thus, in this scaling, the long-time evolution is indeed diffusive. Since the initial value problem does not involve a complicated half-range analysis, connecting the diffusion solution with initial conditions should pose fewer technical difficulties than the analogous boundary value problem \cite{Malvagi1991,Grad1963}.

\smallskip

While preparing this article, we learned of a preprint \cite{BenDor2022} that also treats bulk-boundary correspondence and long-range effects in active colloid steady states, and arrives at similar conclusions regarding the density and current profiles near a planar wall. Their treatment of interparticle interactions is an interesting and nontrivial extension to our analysis, which is restricted to dilute systems. On the other hand, our work goes beyond theirs in that we calculate exactly the boundary layer in the planar problem, as well as treat more general curvilinear geometries. Moreover, by formulating the microscopic transport equation as a boundary value problem, we provide a systematic procedure for constructing and validating bulk-boundary correspondence principles such as the diffusion approximation in section \ref{sec:method_of_solution}.

\smallskip

{\bf Acknowledgments.}
We acknowledge support from the Brandeis Center for Bioinspired Soft Materials, an NSF MRSEC,  DMR-1420382 (CGW, AB, MFH), NSF-MRSEC-2011486 (AB, MFH) , NSF DMR-1149266 and BSF-2014279 (CGW and AB), and DMR-1855914 (MFH). CGW also acknowledges support from the Heising-Simons Foundation, the Simons Foundation, and National Science Foundation Grant No. NSF PHY-1748958. Computational resources were provided by the NSF through XSEDE computing resources (MCB090163) and the Brandeis HPCC which is partially supported by the Brandeis MRSEC and NSF OAC-1920147. We also thank Granville Sewell for assistance with PDE2D.

\appendix
\section{Numerical methods} \label{appendix:numerical-methods}

For the particle-based simulations, we numerically integrate \eqref{eq:introduction-1} and \eqref{eq:introduction-2} using a stochastic Euler algorithm.
Our simulations are ideal for GPU architectures since the code is trivial to parallelize (particles are non-interacting), and so we wrote a CUDA implementation alongside C++ code for CPUs. In most cases we use timestep $\Delta t = \mathcal{O}(10^{-5} D_r^{-1})$. In the C++ code, however, we use $10 \Delta t$ as timestep far from the boundary, where the solution is slowly varying, which results in a substantial speed-up. We take statistics at intervals of $D_r^{-1}$, beginning at a threshold time of around $10^4 D_r^{-1}$. All simulations use $D_t = 0.001 \,\ell^2 D_r$, except for the ellipse problem from section \ref{sec:elliptic-geometry}, where we use $D_t = 0.0001 \ell^2 D_r$. The reason for a smaller value is to ensure the width $\delta$ of the diffusive boundary layer scales appropriately with the dimensions of the ellipse.

For the circular geometry (section \ref{sec:circular-geometry-d}), we use the particle-surface interaction potential
\begin{align}
V(r,\alpha)&=\frac{\tanh\left(100R-100 \, r \right)+1}{10} \left[\tanh \left(5 \alpha-\frac{5\pi}{2}\right) + \tanh \left( \frac{15\pi}{2}-5\alpha \right) - 2 \right] \nonumber \\
&+ \frac{1}{10} \left(r + \frac{1}{19 (r - R + 1)^{19}} \right) - \frac{191}{190} \left[ 1 - \text{H}(R-r) \right] \label{eq:appendix-circular-continuous-potential}
\end{align}
which is continuous in the region of interest. Here $H(z)$ is the Heaviside function. Roughly speaking, the potential gives rise to a ``ratchet'' along the surface of the circle. Where $x < 0$ particles experience a strong but short-ranged force in the direction of $-\hat{r}$. By contrast, fixing $r$ and varying $\alpha$ from the left hemisphere to the right, particles experience a weaker but longer-range force in the direction of $-\hat{\alpha}$, i.e., pointing towards the left hemisphere. The asymmetry between the radial and angular components forms a ratchet that rectifies the motion of the ABPs in agreement with previous work on self-propelled particles in ratchet potentials \cite{Reichhardt2017}. The net effect is that particles tend to be bound on the left side of the sphere and released on the right.

The interactions for the elliptic geometry are defined as follows. For the repulsive force, $\hat{e}_{\eta} \cdot \boldsymbol{F}_r = 0$ and
{ \everymath={\displaystyle}
\begin{align}
\frac{\hat{e}_{\mu} \cdot \boldsymbol{F}_r}{\ell D_r} =
\left\{
\begin{array}{cc}
(10 s)^{-40} - 1 & \qquad  0 < s \leq 0.1  \\
 0 & \qquad s > 0.1
\end{array}
\right. \label{eq:appendix-numerics-3}
\end{align}
}
where $s$ is the distance from an ellipse with semiminor and semimajor axes $a$ and $b$. While we would like to identify effective semimajor and semiminor axes induced by $\boldsymbol{F}_r$, these are not well-defined because the constant-distance surfaces, along which $\hat{e}_{\mu} \cdot \boldsymbol{F}_r$ is constant, are not ellipses. However, since the range of the force is small compared with the major- and minor-axes, this discrepancy is small. Indeed, a numerical calculation shows that effective values $a_{\text{eff}} = a + 0.1 \ell$ and $b_{\text{eff}} = b + 0.1 \ell$ define an ellipse which very nearly coincides with the $s = \ell$ surface.

We use the Fortran-based software PDE2D \cite{Note1} for our finite element computations. For solving 3-dimensional problems like \eqref{eq:introduction-4}, it uses a collocation method with tricubic Hermite basis functions \cite{Sewell2010,Sewell2018}. In our analysis, we use $20$ gridlines in the $\theta$ variable, $50$ in the angular coordinate $\alpha$ or $\eta$, and at least $90$ in the radial coordinate $r$ or $\mu$.

\section{Supplementary material for the 2d planar problem} \label{appendix:2d-problems}

\begin{figure}
\centering
 \includegraphics[keepaspectratio=true,scale=0.4]{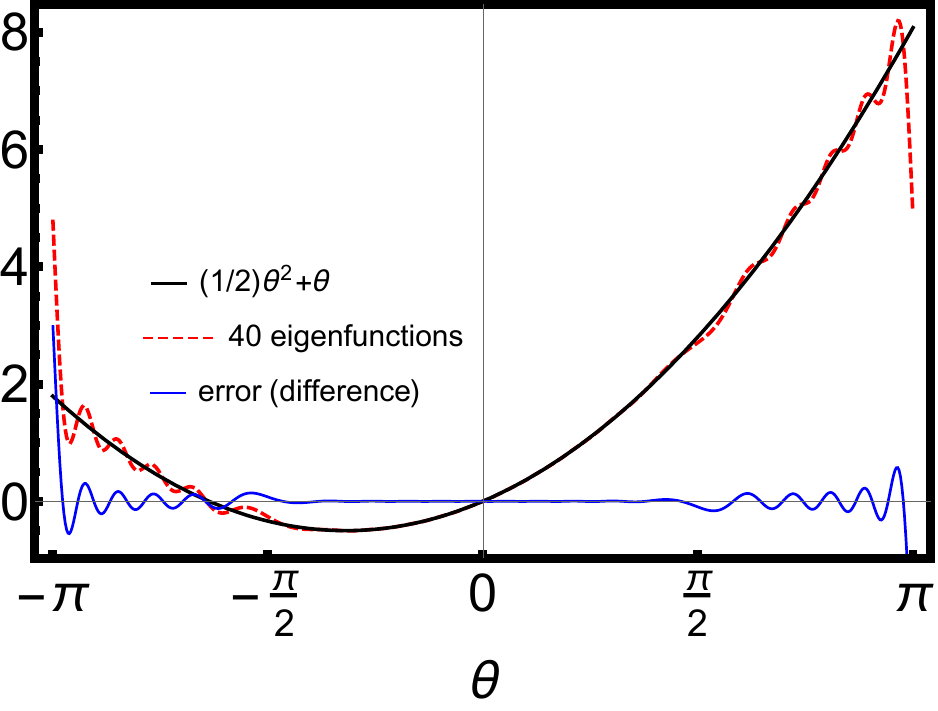}
  \caption[Test of the full-range completeness conjecture for $\mu = 1$]{Test of the full-range completeness conjecture for $\mu = 1$. Using a least-squares technique, we approximate the function $(1/2) \theta^2 + \theta$ (black) as a linear combination of the first 40 eigenfunctions (red dashed). The blue curve shows the difference between the two.}
  \label{fig:full-range-expansion-test}
\end{figure}
\begin{figure}
\centering
 \includegraphics[keepaspectratio=true,scale=0.3]{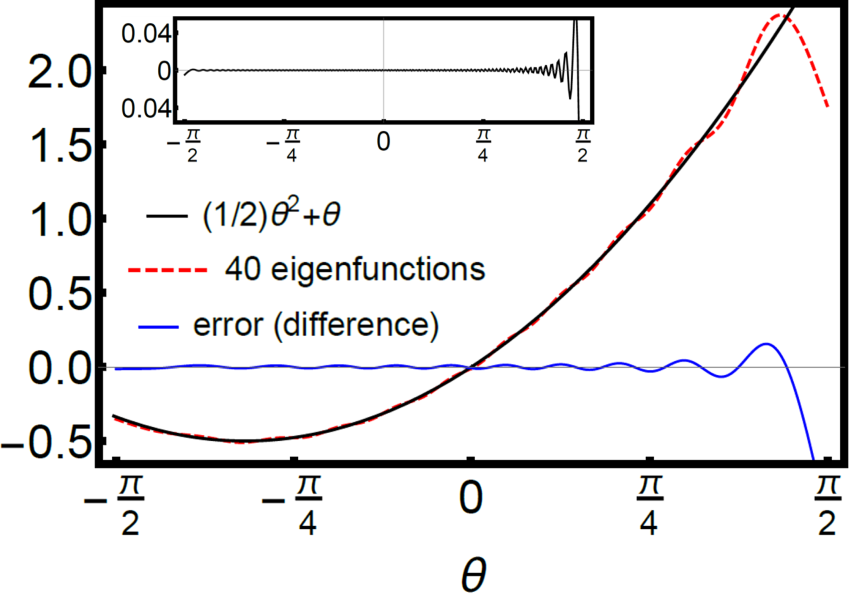}
 \includegraphics[keepaspectratio=true,scale=0.32]{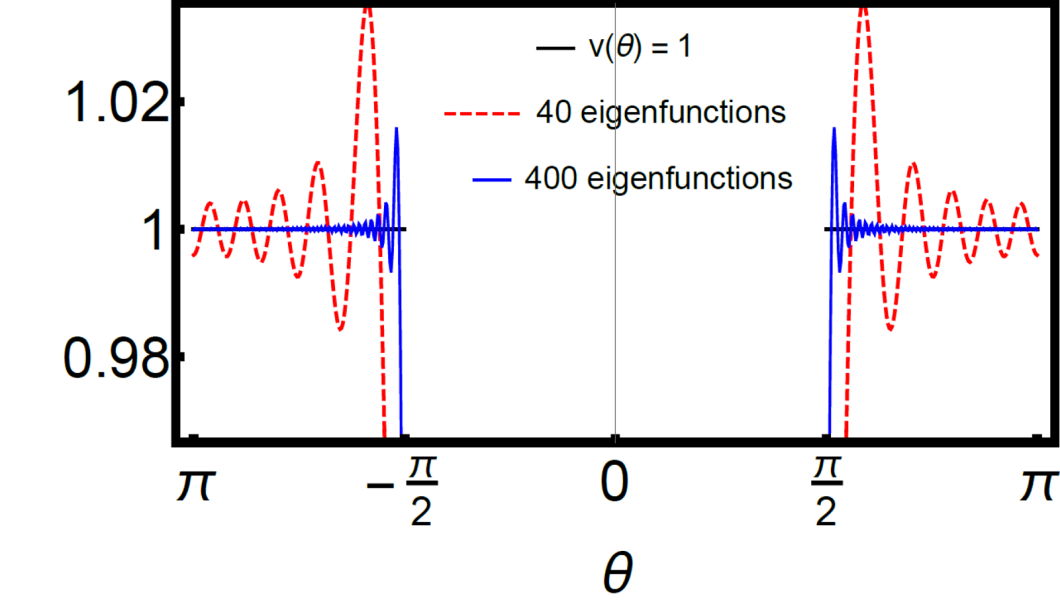}
  \caption[Test of the half-range completeness conjecture for $\mu = 1$]{Test of the half-range completeness conjecture for $\mu = 1$. In the left panel we test the conjecture on the positive half range, $\cos \theta > 0 $. Using a least-squares technique, we approximate the function $(1/2) \theta^2 + \theta$ (black) as a linear combination of the first 40 eigenfunctions (red dashed). The blue curve shows the difference using 40 eigenfunctions, and the inset shows the difference using 400 eigenfunctions. Similarly, in the right panel we test the conjecture on the negative half range, $\cos \theta < 0$, where the function $v(\theta)=1$ is expanded using 40 and 400 eigenfunctions, respectively.}
  \label{fig:half-range-expansion-tests}
\end{figure}

\vspace{2 mm}

\textbf{Completeness conjectures}

\vspace{2 mm}

To test the full-range completeness conjecture, we expand a function $v(\theta)$ in the subspace consisting of the first $2 (N + 1)$ eigenfunctions, for some $N$:
\begin{equation}
v(\theta) = a_0^{+} \Omega_0^{+} +  a_0^{- }\Omega_0^{-} + \sum_{|k| \leq N, k \neq 0} a_k \Omega_k \label{eq:appendix:2d-planar-1}
\end{equation}
where the $\Omega_k$ have arguments $(\theta, |\mu|)$. To determine the coefficients $a_k$, we minimize the squared difference
\begin{equation}
\int_{-\pi}^{\pi} \left[ v(\theta) - a_0^{+} \Omega_0^{+} -  a_0^{- }\Omega_0^{-} - \sum_{|k| \leq N, k \neq 0} a_k \Omega_k \right]^2 |\cos \theta| \, d\theta \label{eq:appendix:2d-planar-2}
\end{equation}
The minimization can be done efficiently on a computer since it only involves the solution of a $2 (N + 1) \times 2 (N+1)$ linear system. The half-range conjectures are tested similarly. For the positive-range conjecture, we minimize
\begin{equation}
\int_{\cos \theta > 0} \left[ v(\theta) - a_0^{+} \Omega_0^{+} - \sum_{0 < k \leq N} a_k \Omega_k \right]^2 |\cos \theta| \, d\theta \label{eq:appendix:2d-planar-3}
\end{equation}
and similarly for the negative-range. 

Representative tests are shown in Figs. \ref{fig:full-range-expansion-test} and \ref{fig:half-range-expansion-tests} -- specifically, taking $v(\theta) = (1/2) \theta^2 + \theta$ on the full range and positive half range, and $v(\theta) = 1$ on the negative half range. All three cases provide good support for the corresponding completeness conjectures.

\vspace{5 mm}

\textbf{Boundary value problem}

\vspace{2 mm}

\begin{figure}
\centering
  \includegraphics[keepaspectratio=true,scale=0.4]{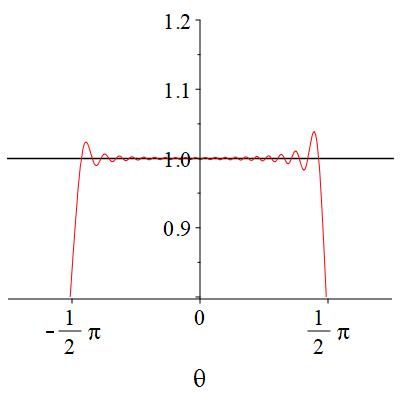} \hspace{10mm}
  \includegraphics[keepaspectratio=true,scale=0.4]{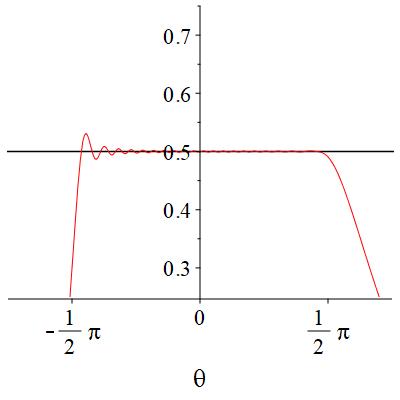}
  \caption[Test of the numerical procedure for calculating the expansion coefficients $a_k, b_k, c_k, d_k$ in Eqs. \eqref{eq:appendix:2d-planar-8}-\eqref{eq:appendix:2d-planar-8}]{Here we test the numerical procedure for calculating the expansion coefficients $a_k, b_k, c_k, d_k$ in Eqs. \eqref{eq:appendix:2d-planar-8}-\eqref{eq:appendix:2d-planar-8} (compare with Eq. \eqref{temp343323543}). We use $\mu = 1$, $q(\theta) = 1$, and $p(\theta) = 1/2$. The left plot shows the l.h.s. of Eq. \eqref{eq:appendix:2d-planar-5} using the calculated $a_k, b_k, c_k, d_k$ and an upper cutoff $k = 40$, and similarly the right plot shows the l.h.s. of \eqref{eq:appendix:2d-planar-6}. We found the l.h.s. of \eqref{eq:appendix:2d-planar-7} and \eqref{eq:appendix:2d-planar-8} to be $0$ within machine precision.}
  \label{fig:bvp-demo}
\end{figure}

Solving the full boundary value problem requires matching the coefficients in the general solution \eqref{temp343323543} to the boundary data. Specifically, if 
\begin{equation}
f(0, y, \theta) = g(y, \theta) = \int_0^{\infty} d\mu \, q(\mu, \theta) \cos \mu y + \int_0^{\infty} d\mu \, p(\mu, \theta) \sin \mu y, \qquad \cos \theta > 0 \label{eq:appendix:2d-planar-4}
\end{equation}
then, for given $\mu$, we must find real constants $a_k, b_k, c_k, d_k$ such that
\begin{align}
&\sum_{k \geq 0} \left(+a_k \Omega_k - b_k \Upsilon_k + c_k \Omega_k + d_k \Upsilon_k \right) = q(\mu, \theta) \qquad \text{where } \cos \theta > 0 \label{eq:appendix:2d-planar-5} \\
&\sum_{k \geq 0} \left(+a_k \Omega_k + b_k \Upsilon_k + c_k \Omega_k - d_k \Upsilon_k \right)= p(\mu, \theta) \qquad \text{where } \cos \theta > 0 \label{eq:appendix:2d-planar-6} \\
&\sum_{k \geq 0} \left(+a_k \Omega_k + b_k \Upsilon_k - c_k \Omega_k + d_k \Upsilon_k \right)= 0 \label{eq:appendix:2d-planar-7} \\
&\sum_{k \geq 0} \left(-a_k \Omega_k + b_k \Upsilon_k + c_k \Omega_k + d_k \Upsilon_k \right)= 0 \label{eq:appendix:2d-planar-8}
\end{align}
where the $\Omega_k$ and $\Upsilon_k$ have arguments $\theta$, $|\mu|$. To solve these equations, we assume the half-range completeness conjecture on the $\Omega_k$. Then, we require that the difference between the l.h.s. and r.h.s. of each equation is orthogonal to each of the $\Omega_j$ with respect to inner product $\langle f, g \rangle = \int_{\cos \theta > 0} f \, g \cos \theta \, d \theta$. In practice, we choose some upper cutoff $k = N$. Then, we obtain the $4 N \times 4 N$ linear system
\begin{equation}
\left(
\begin{array}{cccc}
A & -B & A & B  \\
B & A & B & -A   \\
B & A & -B & A   \\
-A & B & A & B
\end{array}
\right) \left(
\begin{array}{c}
\boldsymbol{\mathrm{a}} \\
\boldsymbol{\mathrm{b}} \\
\boldsymbol{\mathrm{c}} \\
\boldsymbol{\mathrm{d}}
\end{array} \right)
= \left(
\begin{array}{c}
\boldsymbol{\mathrm{u}} \\
\boldsymbol{\mathrm{v}} \\
0 \\
0
\end{array}
\right) \label{eq:appendix:2d-planar-9}
\end{equation}
where $\boldsymbol{\mathrm{a}}, \boldsymbol{\mathrm{b}}, \boldsymbol{\mathrm{c}}, \boldsymbol{\mathrm{d}}$ are $N$-dimensional vectors whose components are the coefficients $a_k, b_k, c_k, d_k$, $k \leq N$; and
\begin{align}
A_{ij} = \langle \Omega_i, \Omega_j \rangle&; \qquad B_{ij} = \langle \Omega_i, \Upsilon_j \rangle \label{eq:appendix:2d-planar-10} \\
\boldsymbol{\mathrm{u}}_i = \langle q, \Omega_i \rangle&; \qquad \boldsymbol{\mathrm{v}}_i = \langle p, \Omega_i \rangle; \label{eq:appendix:2d-planar-11}
\end{align}
We have found $N = 40$ to be sufficient in most cases, although we have tested up to $N = 400$ in some calculations. Fig. \ref{fig:bvp-demo} shows an example with $\mu = 2$, $q(\theta) = 1$, $p(\theta) = 1/2$, and $N = 40$.  

\section{Locally planar approximation} \label{appendix:local-planar}

Here we illustrate the type of ``locally planar" approximation that can be used to estimate the boundary normal flux. We use the same setup as from section \ref{sec:circular-geometry-d}, that is, a circular geometry with boundary data
\begin{align}
f(R, \alpha, \theta) &=
\left\{
\begin{array}{cc}
1  & \quad \text{where } \cos \alpha > 0 \text{ and } \cos (\theta -\alpha) > 0  \\
0 & \quad \text{where } \cos \alpha < 0 \text{ and } \cos (\theta -\alpha) > 0 \\
\text{unspecifed} & \text{else}
\end{array}
\right. \label{eq:appendix:local-planar-1} \\
\lim_{r \rightarrow \infty} f(r, \alpha, \theta) &= c_0 = \text{constant} \label{eq:appendix:local-planar-2}
\end{align}
to be solved on $R < r < \infty$, $0 < \alpha < 2 \pi$, $-\pi < \theta < \pi$. The constant $c_0$ is constrained by the boundary condition at $r = R$. In a planar problem, $c_0$ would correspond to the constant Fourier mode of \eqref{eq:appendix:local-planar-1}, which in the present case gives $c_0 = 1/2$. For the range of $R$-values considered here, the actual value is slightly smaller and carries a weak $R$-dependence. For instance, $c_0 \approx 0.466$ for $R=1$ and $0.452$ for $R=5$. However, consistent with the level of approximation adopted elsewhere in this section, we simply take $c_0 = 1/2$. Then, it is convenient to subtract $c_0$ from the overall solution and solve the transformed problem
\begin{align}
f(R, \alpha, \theta) &=
\left\{
\begin{array}{cc}
+1/2  & \quad \text{where } \cos \alpha > 0 \text{ and } \cos (\theta -\alpha) > 0  \\
-1/2 & \quad \text{where } \cos \alpha < 0 \text{ and } \cos (\theta -\alpha) > 0 \\
\text{unspecifed} & \text{else}
\end{array}
\right. \label{eq:appendix:local-planar-3} \\
\lim_{r \rightarrow \infty} f(r, \alpha, \theta) &= 0 \label{eq:appendix:local-planar-4}
\end{align}

\begin{figure}
\centering
  \includegraphics[keepaspectratio=true,scale=0.37]{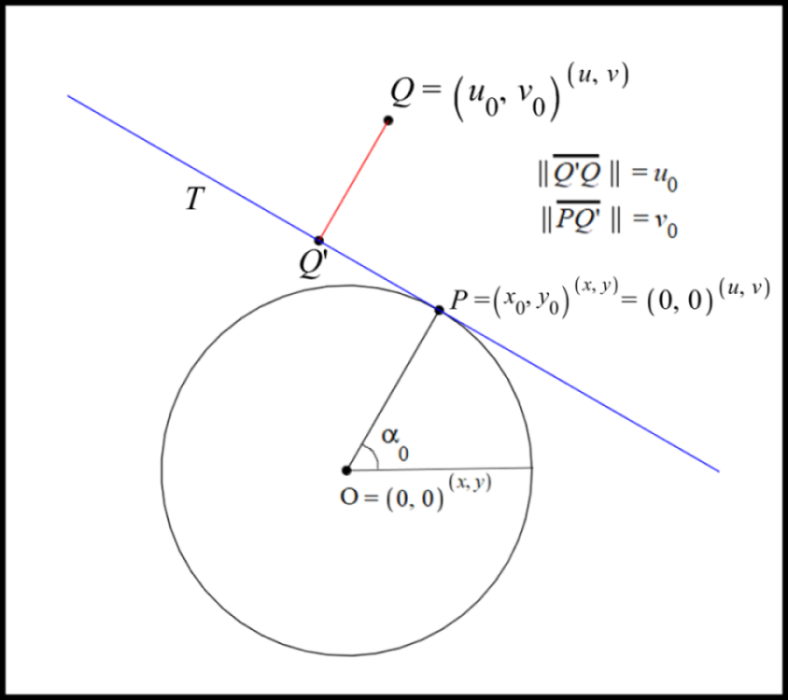}
  \includegraphics[keepaspectratio=true,scale=0.37]{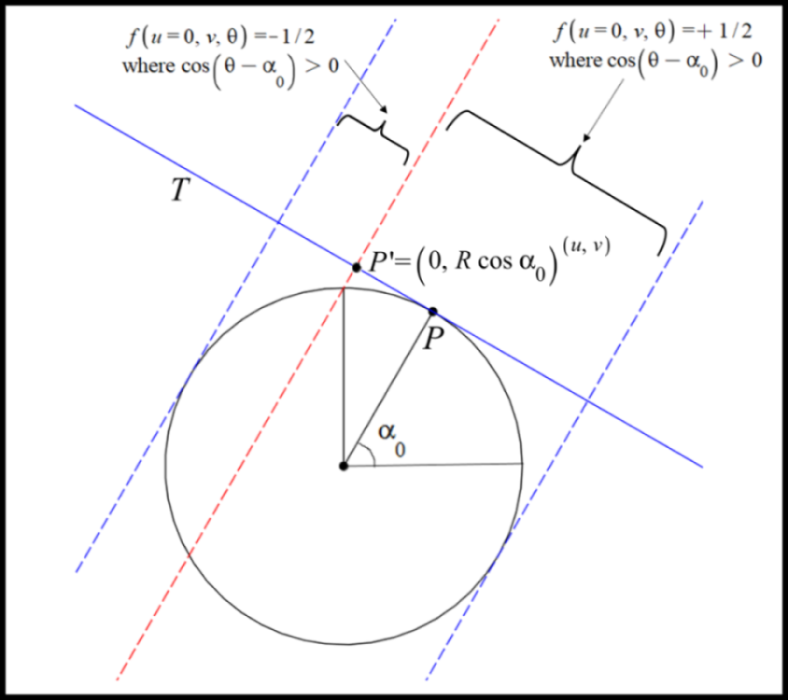}
  \caption[Ingredients in the locally planar approximation]{Ingredients in the locally planar approximation. The left panel illustrates the $(u, v)$ coordinate system corresponding to a point $P$ on the boundary. As discussed in appendix \ref{appendix:local-planar}, $u$ and $v$ are the horizontal and vertical coordinates in a planar half-space problem bounded on the left by the tangent line $T$ ($u = 0$). The right panel shows how we approximate the boundary conditions at $u = 0$ by projecting onto $T$ the boundary conditions \eqref{eq:appendix:local-planar-5}, which sit on the surface of the sphere. The idea is that particles retain their approximate angular distribution as they leave the sphere and pass through $T$.}
  \label{fig:locally_planar_geometry}
\end{figure}

The locally planar approximation is illustrated in Fig. \ref{fig:locally_planar_geometry}. Consider a point $P$ on the boundary, with Cartesian coordinates $(x_0, y_0)$ and polar coordinates $(R, \alpha_0)$, and tangent line $T$. We can define an infinite half-space problem if we consider $T$ as defining a planar boundary: if $f(r, \alpha, \theta)$ is known along the entirety of $T$, then we can use the methods of section \ref{sec:planar-geometry} to solve for $f(r, \alpha, \theta)$ on the infinite half space outside of $T$ (on the side opposite the circle). 

Towards this end, for a point $Q$ above $T$, let $u$ be the perpendicular distance from $T$ and $v$ be the parallel distance from $P$. We denote these coordinates with superscript $(u,v)$ such that $Q = (u,v)^{(u,v)}$ and $P = (0,0)^{(u,v)}$. In terms of these coordinates, suppose we know $f(u = 0, v, \theta)$ on $-\infty < v < \infty$ and where $\cos (\theta - \alpha_0) > 0$. Then, $f(u,v,\theta)$ solves an infinite half-space problem on $u>0$ of the type solved analytically in section \ref{sec:planar-geometry}, with boundary conditions $f(u = 0, v, \theta) = g(v, \theta)$ defined on $-\infty < v < \infty$, $\cos (\theta - \alpha_0) > 0$. It remains to determine $g(v, \theta)$. 

Suppose we can satisfy ourselves with solving for $f$ only very close to $P$. This is certainly true if our goal is to determine the boundary normal flux at $P$. Being close to $P$ means that the form of $g(u,\theta)$ for large $|v|$ influences the solution much less than its behavior near $v= 0$, so we can get away with some approximations in the former region. In this spirit, the locally planar approximation projects the boundary conditions at $r = R$ onto $T$. For the point $P$, for instance, we would take
\begin{align}
\left. g(v, \theta) \right|_{\cos (\theta - \alpha_0) > 0} &= \left\{
\begin{array}{cc}
+1/2  & \quad -R < u < R \cos \alpha_0  \\
-1/2 & \quad R \cos \alpha_0 < u < R \\
0 & |u| > R
\end{array}
\right. \label{eq:appendix:local-planar-5} \\
\lim_{r \rightarrow \infty} f(r, \alpha, \theta) &= a_0 = \text{constant} \label{eq:appendix:local-planar-6}
\end{align}
Though less accurate for large $|v|$, this approximation will generate accurate boundary conditions near $P$, and the exact boundary condition at $P$ itself. Thus, resulting solution $f(r, \alpha, \theta)$ is also expected to be accurate near $P$.

\begin{figure}
\centering
  \includegraphics[keepaspectratio=true,scale=0.4]{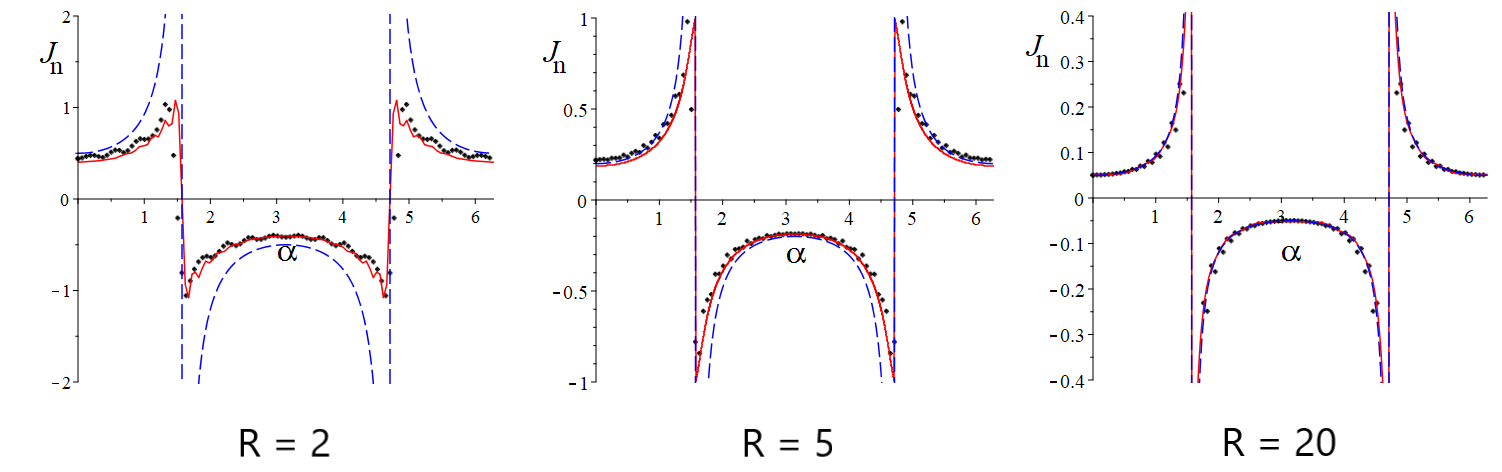}
  \caption{Some tests of the locally planar approximation for $R = 2, 5, 20$. The black diamonds show the normal flux $J_{\text{n}}(\alpha)$ evaluated numerically using PDE2D. The solid red curve shows the locally planar approximation described in appendix \ref{appendix:local-planar}, and the blue dashed curve shows the function $(R \cos \alpha)^{-1}$. The accuracy of the approximations increases with $R$, but is good even for $R = 2$.} \label{fig:locally_planar_test}
\end{figure}

Fig. \ref{fig:locally_planar_test} tests the approximation for several $R$, comparing the calculated boundary normal flux $J_{\text{n}}(R, \alpha)$ with the finite-element solution. The agreement is surprisingly good over a range of $R$ values. If $R$ is large and $\cos \alpha$ not too close to $0$, then $J_{\text{n}}(R, \alpha)$ is $J_{\text{n}}(R, \alpha) \approx \left(R \cos \alpha\right)^{-1}$, which is also shown in Fig. \ref{fig:locally_planar_test}.

\bibliographystyle{apsrev4-1}
\bibliography{bib}

\end{document}